\documentclass[reprint,
amsmath,
amssymb,
aps,
prb,
groupedaddress,
nofootinbib,
nobibnotes,
 floatfix, 
 superscriptaddress,
 longbibliography]{revtex4-1}

\usepackage{times}
\usepackage{amsmath, amssymb, amsthm}
\usepackage{blkarray}
\usepackage{graphicx}
\usepackage{bm,bbm}
\usepackage{hyperref}
\usepackage[margin=2cm]{geometry}
\usepackage{xcolor}
\usepackage{algorithm}
\usepackage[noend]{algpseudocode}
\usepackage{multirow}
\usepackage{url}
\usepackage{mathtools}
\usepackage{float}
\usepackage{dcolumn}
\usepackage{amsmath}
\usepackage{verbatim}

\begin{document}

\title{Fermionic Wave Functions from Neural-Network Constrained Hidden States}

\author{Javier Robledo Moreno}
\email{jrm874@nyu.edu}
\affiliation{
Center for Computational Quantum Physics, Flatiron Institute, New York, NY 10010 USA
}
\affiliation{Center for Quantum Phenomena, Department of Physics, New York University, 726 Broadway, New York, New York 10003, USA
}

\author{Giuseppe Carleo}
\email{giuseppe.carleo@epfl.ch}
\affiliation{
Institute of Physics, \'{E}cole Polytechnique F\'{e}d\'{e}rale de Lausanne (EPFL), CH-1015 Lausanne, Switzerland \\
}

\author{Antoine Georges}
\email{ageorges@flatironinstitute.org}
\affiliation{Collège de France, 11 place Marcelin Berthelot, 75005 Paris, France}
\affiliation{Center for Computational Quantum Physics, Flatiron Institute, New York, NY 10010 USA}
\affiliation{Centre de Physique Théorique, Ecole Polytechnique, CNRS, 91128 Palaiseau Cedex, France}
\affiliation{Department of Quantum Matter Physics, University of Geneva, 24 Quai Ernest-Ansermet, 1211 Geneva 4, Switzerland}

\author{James Stokes}
\email{jstokes@flatironinstitute.org}
\affiliation{
Center for Computational Quantum Physics, Flatiron Institute, New York, NY 10010 USA
}
\affiliation{
Center for Computational Mathematics, Flatiron Institute, New York, NY 10010 USA}

\date{\today}

\begin{abstract}
We introduce a systematically improvable family of variational wave functions for the simulation of strongly correlated fermionic systems. 
This family consists of Slater determinants in an augmented Hilbert space involving `hidden' additional fermionic degrees of freedom. 
These determinants are projected onto the physical Hilbert space through a constraint which is optimized, together with the single-particle 
orbitals, using a neural network parametrization. 
This construction draws inspiration from the success of hidden particle 
representations~\cite{note_slave} but overcomes the limitations associated with the mean-field treatment of the constraint often used in this context. 
Our construction provides an extremely expressive family of wave functions, which is proven to be universal. 
We apply this construction to the ground state properties of the Hubbard model on the square lattice, 
achieving levels of accuracy which are competitive with \textit{state-of-the-art}  variational methods.
\end{abstract}

\maketitle

Many-body quantum systems are computationally challenging because of the exponential dependence of the size of the Hilbert space 
on the number of particles. Variational approaches address this problem by considering a class of wave functions depending on a 
set of parameters over which an optimization is performed.
In this way, the computationally intractable search over the full Hilbert space is reduced to a search over a submanifold of dimension merely polynomial in the number of particles. 
Variational approaches have proven successful in providing qualitative and quantitative insights  
into the nature of the ground state and the low-energy excited states of a number of interacting quantum systems.
For example, in the case of spin systems with arbitrary pairwise interactions, it has been proven~\cite{lieb1973classical, bravyi2019approximation} that the ratio between the energy of optimized mean-field states and the true ground state energy approaches a finite constant in the limit of large system size. The subsequent development of systematically improvable variational wave functions has led to quantitative agreement with exact energies of one-dimensional systems using matrix product states, and recently also in two dimensions using neural-network and tensor-network states~\cite{hyatt20202D_tensor_nets, Choo2019J1J2}. 

The remarkable 
success of variational states in the description of quantum spin systems unfortunately does 
not have a parallel in correlated systems of fermions, however. 
It is known, for example, that the natural mean-field analogue of direct-product states, the so-called Slater determinant (SD) states, fail to even qualitatively describe the thermodynamic limit of Fermi-Hubbard type Hamiltonians~\cite{bravyi2019approximation}  and the development of systematically improvable neural-network-based trial wave functions is currently an active field of research both in second quantization~\cite{ Choo2020chemistry, Yoshioka2021solids, bennewitz2021VMC+VQE}, and first quantization~\cite{Nomura2017slaterRBM, Luo2019backflow, Stokes2020, Pfau2020Ferminet, Spencer2020better, Hermann2020Paulinet, Inui2021determinantfree}. In the latter approach, the wave function amplitudes must be anti-symmetric functions of the particle configurations, while being able to capture correlations beyond the single-particle Slater determinants. This is typically achieved either by considering determinants of multi-particle orbitals~\cite{Luo2019backflow, Pfau2020Ferminet, Hermann2020Paulinet} (backflow transformations), or by Slater determinants of single-particle orbitals multiplied by a neural-network Jastrow factor that depends on the lattice occupations~\cite{Nomura2017slaterRBM, Stokes2020}.  Despite being universal in the lattice, the Slater neural-network Jastrow wave functions seem to struggle to get competitive energies in the strong coupling regime.

The Hubbard model on the square lattice has been the subject of intense theoretical scrutiny, 
and constitutes the most iconic `simple' model of an interacting quantum system. Despite this 
simplicity, a full computational solution is still to be achieved. 
For this model, as well as related lattice models of interacting fermions such as the $t$-$J$ and Kondo lattice models, 
significant insight
has been obtained using hidden particle approaches~\cite{note_slave}. 
Although a number of different formulations are available
~\cite{barnes_1976,barnes_1977,coleman_1984,
Kotliar1986SlaveBosons,kotliar_largeN,
li_rotinv_1989,
fresard_1992,slavefermion1,slavefermion2,
Florens2002SlaveRotor,Florens2004Slaverotor,Medici2005slavespins,lechermann_2007, 
lanata_ghostGA_2017,frank_ghost_2021,guerci_2019,guerci_phd_2019,ancilla_zhang_2020,ancilla_nikolaenko_2021}
all such approaches share a basic concept which 
consists in augmenting the physical Hilbert space by auxiliary degrees of freedom and subsequently performing a 
projection back to the subspace of physical states.
This projection can be regarded as a constraint that selects the representative states in the augmented space that are 
identified with the basis of the physical Hilbert space. 
In many cases, a mean-field saddle point approximation is 
applied both to the auxiliary particle hamiltonian and to the treatment of the constraint, which is implemented 
with static and uniform Lagrange multipliers. 
This mean-field approximation is uncontrolled in general, except when the saddle-point is associated with 
the limit of large number of flavors~\cite{coleman_1984,kotliar_largeN}. Even in those cases, going beyond the saddle point 
level is challenging and no systematic improvements beyond the mean-field variational wave functions are available, 
especially in view of the approximate treatment of the constraint. 

In this article, we draw inspiration from hidden particle approaches to construct 
a systematically improvable family of variational fermionic wave functions. 
These states are obtained as the {\it exact} projection of Slater determinant states 
in a Hilbert space augmented by hidden fermionic degrees of freedom. 
One of the major novelties of the proposed method is that the constraint is parametrized
%
by neural networks, giving rise to an extremely flexible family of wave function {\it ans\"{a}tze}. The constraint is optimized together with the orbitals in the enlarged Hilbert space with the goal of minimizing the energy. The expressive power of this new class of wave functions is demonstrated in a variational Monte Carlo (VMC) setting, 
obtaining an accuracy which is competitive with the  \textit{state-of-the-art} for the ground-state properties of the Hubbard model in the square and rectangular lattices.

The paper is structured as follows: we begin (Sec.~\ref{sec_background}) 
by introducing the Hamiltonian and the physical degrees of freedom of the problem. 
In Sec.~\ref{sec_formalism}, we introduce the fundamentals of the hidden fermion representation, describe the Slater determinant in the augmented space together with the fully parametrized constraint function 
and prove the universality of this representation. This section also contains details on the VMC implementation. 
In Sec.~\ref{sec_experiments}, 
we present ground-state energy benchmarks for the Hubbard model with increasingly large system sizes, 
and demonstrate that we can stabilize competing orders of charge and spin stripes for the Hubbard model on rectangular lattice geometries.

\section{Background: States and Hamiltonian}
\label{sec_background}

In this paper we develop a general technique for approximating the ground state of interacting fermionic hamiltonians with discrete degrees
of freedom--as defined for example by discrete orbitals or spatial coordinates. 
As a specific application, we focus here on the Fermi-Hubbard model, whose Hamiltonian reads 
\begin{equation}\label{eq: Hubbard}
    \hat{H} = -\sum_{\sigma \in \{ \uparrow, \downarrow \}}\sum_{\{i,j\}\in\mathcal{E}}t_{ij}\big( \hat{c}_{i\sigma}^\dag \hat{c}_{j\sigma} + \hat{c}_{j\sigma }^\dag \hat{c}_{i\sigma}\big) + \sum_{i \in \mathcal{V}} U_i \hat{n}_{i\uparrow} \hat{n}_{i\downarrow} \enspace ,
\end{equation}
where the binary index $\sigma \in \{\uparrow,\downarrow \}$ labels two species of fermionic modes satisfying the canonical anti-commutation relations,
\begin{equation}
    \{\hat{c}_{i\sigma}^\dag, \hat{c}_{j\sigma'} \} = \delta_{ij}\delta_{\sigma\sigma'} \enspace ,\quad \quad \{\hat{c}_{i\sigma}, \hat{c}_{j\sigma'} \} = 0 \enspace .
\end{equation}
The fermionic modes $\hat{c}_{i\sigma}$ are the \textit{physical} (electronic) degrees of freedom (DOF). The fermion dynamics is described by the lattice with $\mathcal{V}$ sites defined by the non-zero entries of the $t_{ij}$ hopping matrix, as well as by the onsite coulomb repulsion $U_i$. In the following, we exclusively focus on the square and rectangular lattices with uniform hopping ($t_{ij} = 1$) and onsite repulsion, leaving more general geometries to future studies. 


In this work we are concerned with the subspace of definite particle numbers $N_\uparrow$ and $N_\downarrow$ of the individual spin species, in which case the two species are distinguishable from each other. However, it is convenient to impose full anti-symmetry between the spin species to enhance the expressivity of the family of trial wave functions, like in the so called unrestricted Hartree-Fock (HF). The projection to definite $N_\uparrow$ and $N_\downarrow$ subspace is imposed in the sampling of the wave function amplitudes.
\begin{figure}
    \centering
            \includegraphics[width=.99\linewidth]{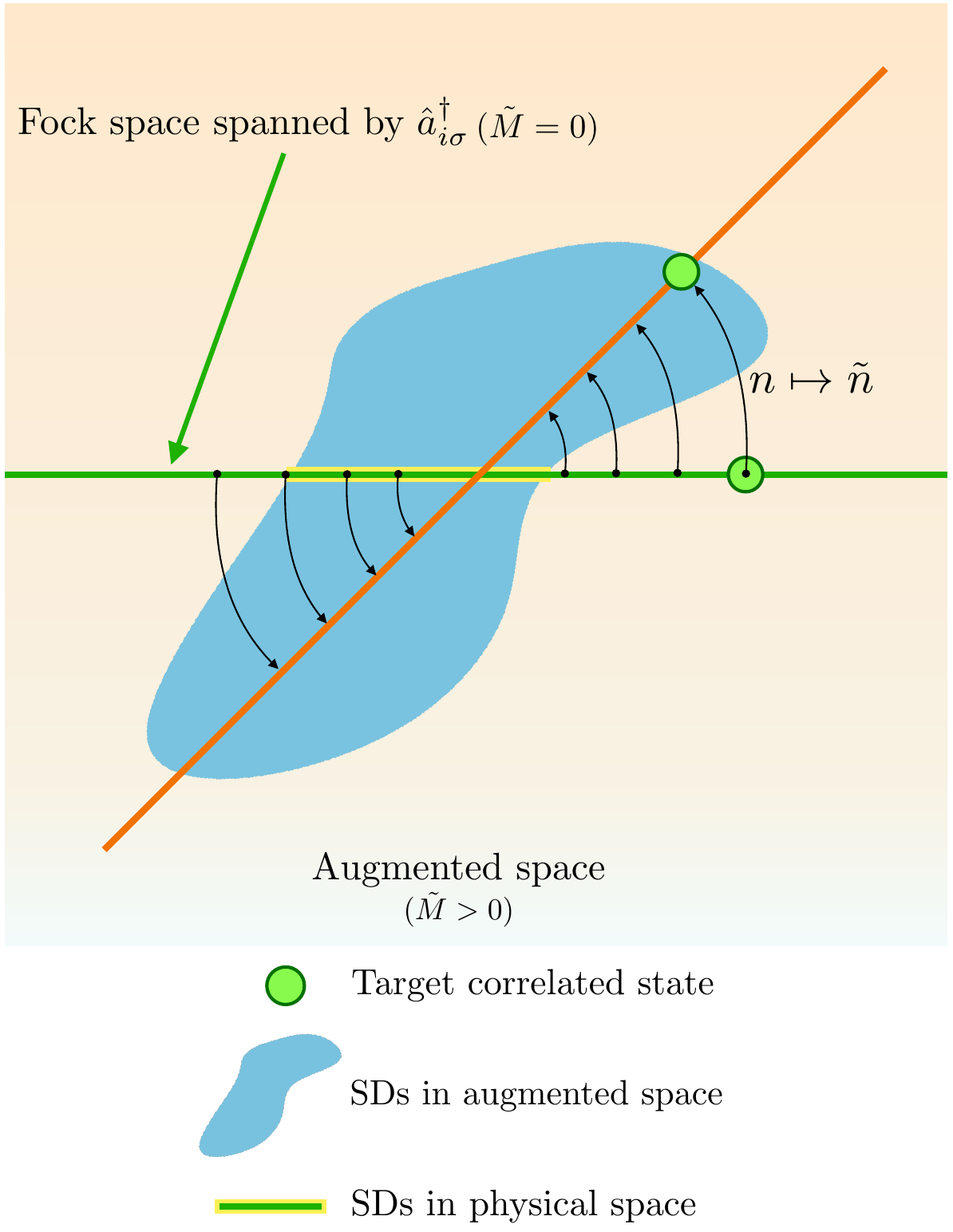}
            \caption{\label{fig_01: geometrical interpretation} Depiction of the geometrical interpretation of the hidden fermion formalism. The Fock space spanned by the visible fermionic modes $\hat{a}^\dag_{i\sigma}$ is represented by the green horizontal line.  The augmented Fock space is represented by the light orange plane (plane of the paper). The orange diagonal line represents the subspace in the augmented Fock space that is isomorphic to the physical Hilbert space after applying the constraint function (black arrows). The collection of Slater determinants (SDs) in the augmented space is represented by the blue shape, and the intersection with the subspace of just visible DOFs is marked in yellow. This intersection corresponds to the physical Hartree-Fock states. The constraint function changes the collection of states that represent the physical Hilbert space bringing the target correlated state close to a Slater determinant in the enlarged space.}
\end{figure} 

\begin{figure*}
            \includegraphics[width=.9\linewidth]{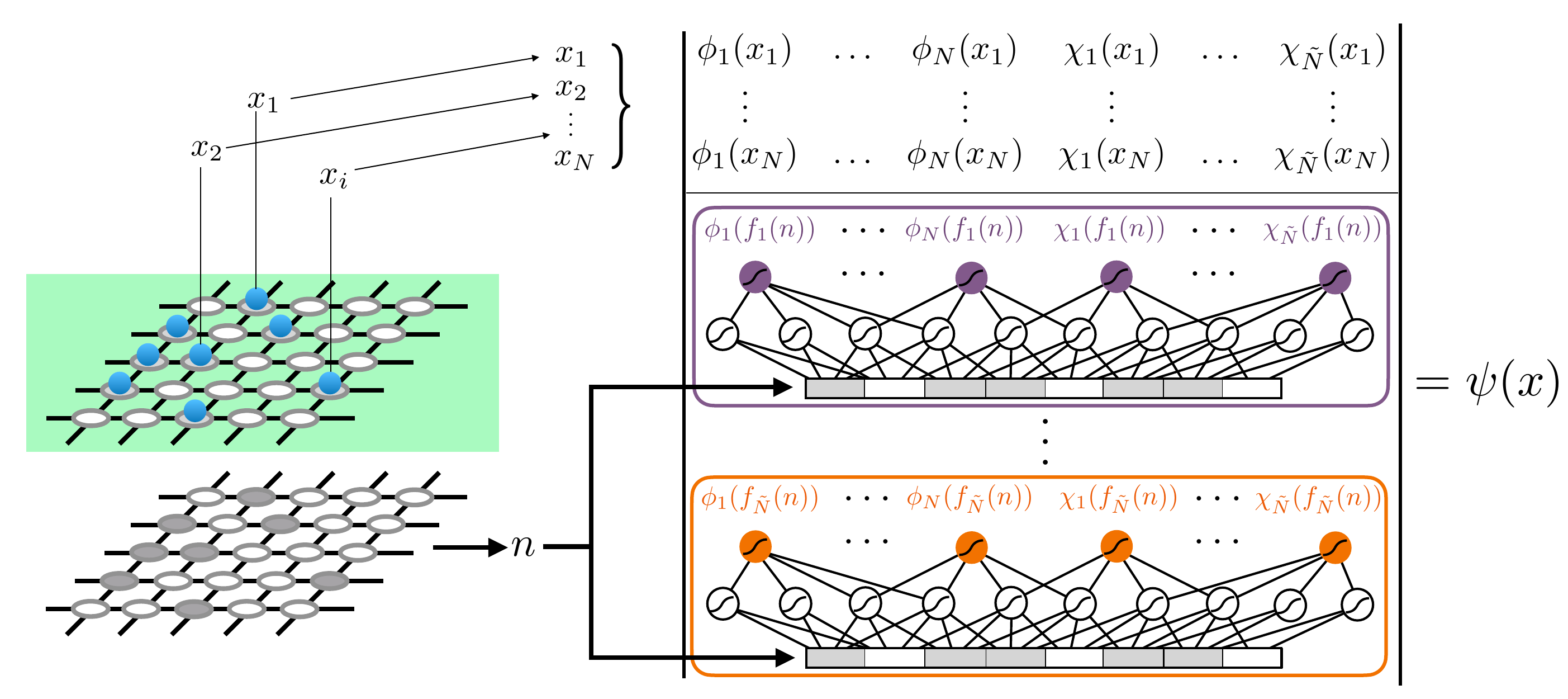}
            \caption{\label{fig_02: hidden fermion SD} Hidden-fermion determinant state amplitudes with a neural-network parametrized constraint function. The top part of the determinant is constructed by slicing $N$ rows from the top $M$ rows of the $\Phi$ matrix, according to visible particle configuration $x$. Each row of the bottom sub-matrix $\big[\phi_{\rm h}(f(x)), \chi_{\rm h}(f(x))\big]$ (hidden sub-matrix) is parametrized by the outputs of a separate neural network (indicated by different colors), whose input is the flattened visible-lattice occupancy $n$.}
\end{figure*}

\section{Hidden fermion formalism and wave function {\it ansatz}}
\label{sec_formalism}

\subsection{States in the augmented Hilbert space: constraint function}

Recall that the multi-particle physical Hilbert (Fock) space is spanned by $M := 2 \mathcal{V}$ creation operators $\hat{c}_{i\sigma}^\dag$ applied in all possible ways to the Fock vacuum $|0\rangle$. The strategy of this paper is to define an augmented Fock space, constructed by $M_{\rm tot}>M$ fermionic modes. 

We partition the mode operators of the augmented Fock space into two species of 
auxiliary fermionic degrees of freedom $\hat{a}^\dag_\mu$ and $\hat{d}^\dag_\nu$, referred to as visible and hidden modes respectively. We note that, although most hidden particle approaches enlarge the Hilbert space with bosonic 
degrees of freedom, fermionic hidden sectors have been considered in recent works~\cite{lanata_ghostGA_2017,frank_ghost_2021,guerci_phd_2019,guerci_2019,ancilla_zhang_2020,ancilla_nikolaenko_2021} 
(see also~\cite{slavefermion1,slavefermion2,slavefermion3}). We require  $1\leq \mu \leq M$, and $1\leq \nu \leq \tilde{M}$ with $\tilde{M}$ a free hyper-parameter. Of course $M_{\rm tot} = M + \tilde{M}$. The occupancy of the visible modes $\hat{a}^\dag_\mu$ is identified one-to-one with the occupancy of the physical modes $\hat{c}_{i\sigma}^\dag$, establishing a direct correspondence between the index $\mu$ of the visible modes and the position-spin multi-index of the physical modes $(i, \sigma)$. 

Thus, the basis for the augmented Fock space is spanned by the set of states
    \begin{equation}\label{eq: augmented Fock basis}
        \left|n, \tilde{n} \right\rangle = \left(\prod_{i,\sigma} (\hat{a}_{i\sigma}^\dag)^{n_{i\sigma}}\right) \left(\prod_{\mu=1}^{\tilde{M}} (\hat{d}_\mu^\dag)^{\tilde{n}_\mu}\right) |0 \rangle,
    \end{equation}
where $n$ and $\tilde{n}$ label the occupancy of the visible and hidden modes respectively. Note that this basis does not have a definite number of hidden fermions, even if the visible occupations are constrained to have definite particle number.

Since the augmented Fock space defines a superset of the physical many-body fermionic states, a collection of "representative" states is chosen to span the basis of the physical Hilbert space within the augmented space; similarly to the constraint applied in the hidden rotor, spin or boson formalism~\cite{Florens2002SlaveRotor, Florens2004Slaverotor, Medici2005slavespins, Kotliar1986SlaveBosons}. This choice produces a basis of the correct dimension, eliminating so called \textit{unphysical states}. This constraint is applied by the following procedure: for each visible fermion occupancy $n$ a particular hidden fermion occupancy $\tilde{n}$ is chosen. The arbitrary choice of the population of the hidden modes can be summarized by a constraint function $F(n)=\tilde{n}$. 

In the physical subspace the probability amplitude of the spinful fermion occupancy $n$ is given by the overlap between the augmented basis states (Eq.~\ref{eq: augmented Fock basis}) and a given trial state vector $|\Psi\rangle$ of the augmented Fock space, where the hidden occupancy $\tilde{n}$ is controlled by the visible occupancy $n$ via the constraint:
\begin{equation}
    \psi(n)=\langle n , F(n) | \Psi \rangle.
\end{equation}
In this work we consider the search of the optimal constraint function, contrary to previous hidden-particle formulations where the constraint is a fixed physically-motivated rule. The resulting wave function {\it ansatz} is thus parametrized by both the choice of the state in the augmented space $|\Psi\rangle$ and the constraint function $F(n)$. Since there exist doubly exponentially many constraint functions, an extremely flexible family of correlated trial wave functions is obtained.

It should be noted that the hidden single-particle orbitals $\hat{d}^\dag_{\mu}$ define an abstract space of hidden particle configurations. While one may be concerned by the nature of this abstract space and the form of the orthogonal set of single particle orbitals that define its basis, in practice, we work in the basis of particle configurations in the orbitals $\hat{d}^\dag_{\mu}$. Consequently, the relevant quantity is the combination of single-particle orbitals in the abstract space.

Fig.~\ref{fig_01: geometrical interpretation} illustrates geometrically the general concept of the hidden fermion formalism. The constraint function can be interpreted as a non-trivial rotation of the collection of states that constitute the basis of the physical Fock space embedded in the augmented space (light green horizontal line is rotated to the orange segment in Fig.~\ref{fig_01: geometrical interpretation}). The goal of this transformation is to bring the target correlated state close to the parametrized family of states in the enlarged space. In Fig.~\ref{fig_01: geometrical interpretation} the chosen family of parametrized states is the family of SDs. We also show that in the particular case of $\tilde{M} = 0$ (light green subspace) the physical Fock space is directly spanned by the visible modes, and that standard SD states can be recovered in that limit.

\subsection{Hidden fermion determinant states}
\label{subsec_hiddenfermions}

\subsubsection{Generalities}
In order to demonstrate the versatility of the hidden fermion approach, we consider in this work the special case where $|\Psi\rangle$ is the uncorrelated Slater determinant state $|\Psi_{\rm SD}\rangle$, which is characterized by a total number $N_{\rm tot} \geq N$ of orbital functions $\phi_n : \{1,\ldots M_{\rm tot}\} \to \mathbb{C}$ where $1 \leq n \leq N_{\rm tot}$ and  $\tilde{N} = N_{\rm tot}-N$ is the number of added hidden fermions. In particular, $|\Psi_{\rm SD}\rangle$ is obtained from the Fock vacuum as follows,
\begin{equation}\label{eq: HF definition}
    |\Psi_{\rm SD}\rangle = \hat{\varphi}_1^\dag \cdots \hat{\varphi}_{N_{\rm tot}}^\dag |0 \rangle \enspace ,
\end{equation}
where each $\hat{\varphi}_\alpha^\dag$ is a linear combination of the original creation operators, whose coefficients are determined by the corresponding orbital. In terms of the row-vectors 
\begin{equation}\label{eq: HF change of basis}
    (\hat{\varphi}^\dag_1, \hdots, \hat{\varphi}^\dag_{N_{\rm tot}}) = (\hat{a}^\dag_1,\hdots, \hat{a}^\dag_{M}, \hat{d}^\dag_1,\hdots, \hat{d}^\dag_{\tilde{M}})\; \Phi
\end{equation}
where $\Phi$ is the $M_{\rm tot} \times N_{\rm tot}$ matrix whose columns correspond to the orbital functions. It will be convenient to write the matrix of orbitals in the following block form,
\begin{equation}\label{eq: matrix orbitals}
    \Phi = 
    \begin{bmatrix}
    \phi_{\rm v} & \chi_{\rm v} \\
    \phi_{\rm h} & \chi_{\rm h}
    \end{bmatrix} \enspace .
\end{equation}
Where $\phi_{\rm v}$ is the $M \times N$ matrix representing the amplitudes of the visible orbitals evaluated in the visible modes, $\chi_{\rm v}$ is the $M \times \tilde{N}$ matrix representing the amplitude of the hidden orbitals evaluated in the visible modes, $\phi_{\rm h}$ is the $\tilde{M} \times N$ matrix representing the amplitude of the visible orbitals evaluated in the hidden modes and $\chi_{\rm h}$ is the $\tilde{M} \times \tilde{N}$ matrix representing the amplitude of the hidden orbitals evaluated in the hidden modes. 

Since the SD state is an eigenstate of the total number operator, as are both the visible and hidden sectors, and anticipating that particle configurations are sampled in the VMC framework, we can represent the constraint as a mapping between the visible-particle configuration $x = (x_1, ..., x_{N})$ and the hidden-particle configuration $\tilde{x} = (\tilde{x}_1, ..., \tilde{x}_{N})$
\begin{equation}
    f:x \mapsto \tilde{x}.
\end{equation}
In order to respect the Fermi statistics, it is sufficient to choose the function $f$ to be of bosonic nature; that is, invariant under permutations of the visible configuration. The amplitudes of the wave function {\it ansatz} in the configuration basis are thus given by
 \begin{equation} \label{e:sfamp}
    \psi(x) = \langle x, f(x) |\Psi_{\rm SD} \rangle =
    \det
    \begin{bmatrix}
    \phi_{\rm v}(x)    & \chi_{\rm v}(x) \\
    \phi_{\rm h}(f(x)) & \chi_{\rm h}(f(x))
    \end{bmatrix}
    \enspace
\end{equation}
where $\big[ \phi_{\rm v}(x)$, $\chi_{\rm v}(x)\big ]$ and $\big[ \phi_{\rm h}(f(x)), \chi_{\rm h}(f(x))\big]$ denote the $N\times(N+\tilde{N})$ and $\tilde{N}\times(N+\tilde{N})$ sub-matrices obtained from $\big[\phi_{\rm v}, \chi_{\rm v}\big]$ and $\big[ \phi_{\rm h}, \chi_{\rm h}\big]$ respectively by slicing the row entries corresponding to $x$ and $f(x)$. For convenience we denote the $\big[ \phi_{\rm h}(f(x)), \chi_{\rm h}(f(x))\big]$ matrix as the \textit{hidden sub-matrix}.

\subsubsection{Universality and connection to other wave function {\it ans\"{a}tze}}
\label{subsec_universality}

This {\it ansatz} is universal in the lattice. The proof relies on the ability of the determinant in Eq.~\ref{e:sfamp} to represent a universal lookup table of amplitudes that are matched with the amplitudes of an arbitrary target state. In the particular case of $\phi_{\rm h} = 0$ and $\chi_{\rm v} = 0$ the flexibility of the {\it ansatz} is relied upon $\chi_{\rm h}$, as $\phi_{\rm v}$ leads to amplitudes that correspond to an uncorrelated state in the physical space. It is possible to construct the lookup table for $\tilde{N} \geq 1$, requiring $\tilde{M}$ to grow combinatorially fast with the number of physical fermionic modes. See the Supplementary Information (SI)~\cite{SI} for a detailed discussion. It follows that in the general case where $\phi_{\rm h} \neq 0$ and $\chi_{\rm v} \neq 0$, the determinant in the enlarged Fock space does not inherit the nodes of the $\phi_{\rm v}$ orbitals. 

Our construction bears some similarities with backflow~\cite{Kwon1993Backflow, Kwon1998Backflow, Tocchio2008Backflow, Tocchio2011Backflow}, in which orbitals are taken to be functions of the coordinates of all particles. In contrast to regular backflow, only the restriction of orbitals to hidden states have multi-particle position dependence (see \eqref{e:sfamp}). 


Jastrow-like wave function {\it ans\"{a}tze}  of the form 
\begin{equation}\label{eq: slater generalized jastrow}
    \psi_{\rm J} (x) =  J(n) \det [ \phi_{\rm v}(x)],
\end{equation}
where $J(n)$ is an arbitrary function of the visible lattice occupations, also appear naturally in this formalism. This connection clearly materializes by considering $\tilde{N} = 1$ and $\chi_{\rm v} = \phi_{\rm h} = 0$. In this case the amplitudes of the wave function {\it ansatz} are the product of $\det [ \phi_{\rm v}(x)]$ and a symmetric function of the visible particle configuration $\chi_{\rm h}(f(x))$. Note that this class of wave functions includes the physically motivated Gutzwiller and Jastrow factors, as well as generalized neural-network Jastrow factors~\cite{Stokes2020, Nomura2017slaterRBM} applied to Slater determinants. The constraint function reproducing the Gutzwiller state can be found in the SI~\cite{SI}.

 Configuration-interaction (CI) wave functions are also explicitly connected to the hidden fermion determinant state. Using the Laplace expansion of the determinant in Eq.~\ref{e:sfamp} along its last $\tilde{N}$ rows yields a linear combination of $N$-particle Slater determinants. If $\phi_{\rm V}$ and $\chi_{\rm V}$ are chosen to be the $N$ lowest HF orbitals and the first $\tilde{N}$ virtual orbitals respectively, then a CI wave function is obtained, containing all possible (single to $\tilde{N}$-tuple) excitations to the first $\tilde{N}$ virtual orbitals. See SI~\cite{SI} for the detailed derivation.

\subsubsection{Parametrized constraint function, practical implementation}
\label{subsec_constraint}

In contrast to physically motivated constraint functions, a more general approach involves considering $f(x)$ to belong to a parametrized family, whose parameters are optimized, together with the orbitals, in the energy minimization. The variational Monte Carlo seeks optimal parameters $\theta$ for a variational family of wave functions $\psi_\theta(x)$, which are assumed to be differentiable with respect to $\theta$. Although the requirement of differentiability appears to be in tension with the combinatorial nature of $f(x)$, this obstacle is easily overcome by parametrizing instead the composition of functions $\phi (f(x))$ and $\chi (f(x))$, which appears in the hidden sub-matrix of the enlarged determinant (see \eqref{e:sfamp}). This parametrization is connected to the notion of a continuous set of orthogonal hidden modes $\hat{d}^\dag_\mu$, which accounts to $\tilde{M} \to \infty$. Remarkably, this automatically satisfies the condition that $\tilde{M}$ must grow combinatorially with $M$ for the determinant in Eq.~\ref{e:sfamp} to be a universal lookup table of amplitudes. The hidden fermion configurations $f(x)$ are thus represented by some internal state of the parametric function. However, in practice we are never interested in such internal state. 

Since the hidden sub-matrix is a  matrix-valued function which by construction is a permutation-invariant function of the visible configuration $x$, we choose to represent it by neural networks, taking as an input the visible occupation numbers $n$, without loss of generality. Neural networks are the perfect candidate to reduce the intractable complexity of choosing the optimal constraint, as they define an extremely flexible family of functions. Furthermore, sufficiently large neural networks can represent arbitrary contrainst functions, since they satisfy a universal approximation theorem~\cite{Cybenko1989universal}.
The set of variational parameters $\theta$ of our {\it ansatz} consists on the the matrices $\phi_{\rm v}$ and $\chi_{\rm v}$ together with the weights and biases parametrizing the corresponding neural network. Fig.~\ref{fig_02: hidden fermion SD} details the precise parametrization used in this work. In practice, each row of the hidden sub-matrix is parametrized by its own neural network, as shown in Fig.~\ref{fig_02: hidden fermion SD} by different colored neural-network blocks. We consider multilayer perceptrons with hyperbolic-tangent activations. The hyperparameters of the {\it ansatz} include the neural network architecture as well as the number of added hidden fermions $\tilde{N}$.

\begin{figure*}
            \includegraphics[width=.99\linewidth]{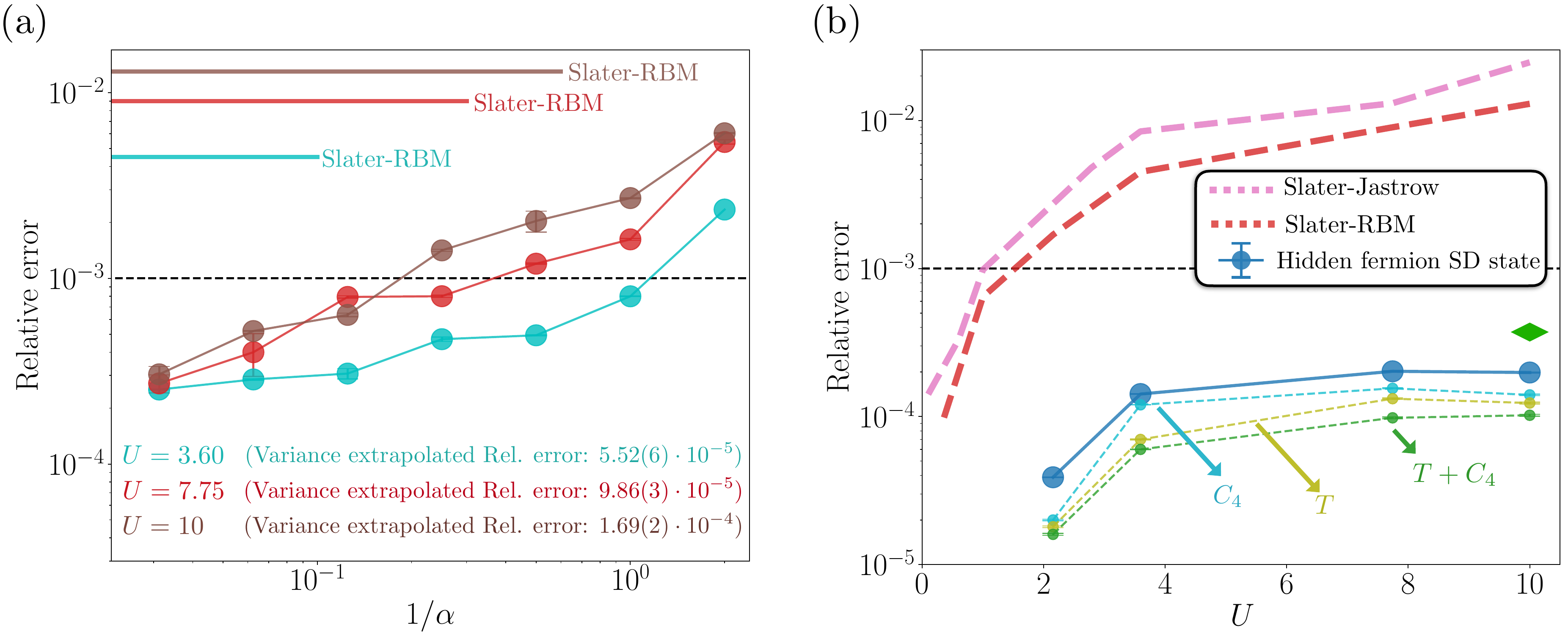}
            \caption{\label{fig_03: 4X4 benchmark} Exact diagonalization benchmarks of the ground-state energy in the $4 \times 4$ lattice with periodic boundary conditions. \textbf{(a)} Relative error in the ground-state energy as a function of the inverse of the width density $\alpha$ of the single-hidden-layer neural networks parametrizing the rows of the hidden sub-matrix. Average physical site occupation is $n = 1/2$ and $\tilde{N} = 8$. Different values of $U$ are considered, as indicated by each color. The error for a Slater-RBM {\it ansatz} (see main text) with hidden neuron density $\alpha = 32$, at the same values of $U$ is included for comparison. Indicated is also the relative error from the variance-extrapolated energy for each value of $U$ (see SI~\cite{SI} for details). \textbf{(b)} Relative error in the ground-state energy as a function of the coupling constant $U$, at $n = 5/8$ average site occupancy (first closed shell) and $\tilde{N} = 10$. The rows of the hidden sub-matrix are given by single-hidden-layer neural networks with $\alpha = 64$. The error from Slater-Jastrow and Slater-RBM {\it ans\"{a}tze}  are included for comparison. The green diamond is the relative error found with the \textit{state-of-the-art}, tensor-network-based {\it ansatz} from Ref.~\cite{Hui-Hai2017treeTN}. Shown is also the relative error according to the projection of the converged hidden fermion determinant state to the subspace of invariant wave functions under the action of $\pi/2$ rotations ($C_4$) and the group of all possible translations $T$ with $K = 0$ momentum, separately and together.}
\end{figure*}

The cost of evaluating the enlarged determinant and its derivatives with respect to the variational parameters scales with the number of visible and hidden fermions as $\mathcal{O}\big( (N + \tilde{N})^3\big)$, coming from the used LU factorization. Typically we choose $\tilde{N}\sim \mathcal{O}\big( N \big)$, and therefore the asymptotic cost of the evaluation of the hidden fermion determinant state is $\mathcal{O}\big( N^3\big)$. The computation of the wave function amplitudes and gradients is the only step in the VMC algorithm where the required resources are larger, by a constant factor, than the computation of a $N$-fermion determinant.

\subsection{Methods}
\label{sec_methods}
Both the amplitudes of the matrices $\phi_{\rm v}$ and $\chi_{\rm v}$, together with the weights and biases of the neural networks parametrizing the rows of the hidden sub-matrix, are jointly optimized using the stochastic reconfiguration method~\cite{sorella2007StochasticReconfiguration}, an extension of the classical natural gradient optimization method~\cite{amari1998natural} to variational quantum states. Given that we are interested in the approximation of the ground-state wave function, we rely on the variational principle and use the expectation value of the Hamiltonian with respect to the variational state as the objective function to be optimized. For every Hamiltonian parameter choice a new trial state is optimized from scratch.

General expectation values and gradients of the objective function are computed using Markov Chain Monte Carlo sampling according to the probability distribution defined by the square of the wave function amplitudes $|\psi_\theta (x)|^2$, working in the basis of particle configurations. We use the Python library NetKet~\cite{netket:2019} for the implementation (see the SI~\cite{SI} for details), where gradients of the wave function amplitudes with respect to the variational parameters are computed by the so called automatic differentiation implemented in the Python library Jax~\cite{jax2018github}.

\section{Numerical Experiments}
\label{sec_experiments}

In this section we benchmark the hidden fermion determinant wave function {\it ansatz} with a fully parametrized constraint function. We first study the square lattice at average site occupation $n = 1/2$ and $n= 5/8$. We use the $4\times 4$ square lattice as a test-bed to study the accuracy (compared to exact diagonalization (ED)) of the proposed {\it ansatz}.  The accuracy is quantified by the difference between the ED ground-state energy and the variational energy, relative to the ED ground-state energy. We analyze the effect of the neural network complexity and compare against relevant results in the literature. Lastly we focus on rectangular geometries of size $4\times L$, where we consider $1/8$ hole-doping ($n = 7/8$). Periodic boundary conditions are set in the short side of the rectangle in all cases. We study the case of both open and periodic boundary conditions on the long side. In the former, we compare our energies with Density Matrix Renormalization Group (DMRG) results and study the competing stripe orders of the system. In the latter and in the smallest system size ($4\times 4$) we analyse the relative error in the ground-state energy, obtained from ED. In the larger sizes ($L = 8$ and $L = 16$) we compare the ground state energy with the results obtained using a Slater-Jastrow {\it ansatz} and the neural network backflow wave function from Ref.~\cite{Luo2019backflow}. In all cases we focus on the zero magnetization and fixed visible and hidden particle subspaces.

\subsection{Benchmarks in the square lattice}
We begin by considering the particular case of $\tilde{N} = N$, which provides a good trade-off between computational complexity and accuracy; and a single-hidden-layer neural network parametrizing each row of the hidden sub-matrix. This architecture is a good starting point to study the effect of the neural network expressive power in the accuracy of the {\it ansatz}. In this case, the expressive power is only determined by the number of hidden units. More hidden units improve the flexibility of the neural network. Furthermore, this single hidden layer architecture is the minimal architecture that satisfies the universal approximation theorem~\cite{Cybenko1989universal}. 

Panel (a) of Fig.~\ref{fig_03: 4X4 benchmark} shows the relative error in the ground state energy as a function of the ratio between the number of hidden units and input features ($\alpha$), at $n = 1/2$ average site occupation. Different values of $U$ are shown, including challenging cases ($U = 7.75$ and $U=10$) where the ground state is strongly correlated. There is a systematic trend to decrease the error in the energy as $\alpha$ is increased, providing a clear and controllable pathway to obtaining more expressive wave function {\it ans\"{a}tze}. Moreover, for the largest values of $\alpha$, and contrary to what is observed on typical wave function {\it ans\"{a}tze}, the error does not significantly increase with $U$ as the correlations in the ground state increase. Remarkably, the relative error at $U = 10$, typically the most challenging case, is orders of magnitude lower than the error of the Slater-RBM wave function {\it ansatz}. The Slater-RBM {\it ansatz} is a particular case of the wave function in Eq.~\ref{eq: slater generalized jastrow}, where $J(n)$ is a restricted Boltzmann Machine of complex weights. 

The direct extrapolation of the relative error to the $\alpha \rightarrow \infty$ limit is challenging as the asymptotic scaling of the accuracy with the neural network complexity is not a well understood matter in the field. However, from the different energy and variance estimates obtained for each $\alpha$ we perform an energy-variance extrapolation procedure~\cite{Kashima2001varianceExtrapolation} to obtain better estimates to the ground-state energy. See SI~\cite{SI} for details. The relative error corresponding to the variance extrapolated energies is shown in Fig.~\ref{fig_03: 4X4 benchmark} (a), where  the relative error is in this case defined as the difference between the variance extrapolated and the ground-state energies, relative to the ground-state energy.

The improvement of the accuracy with the increase of $\alpha$ is accompanied by a gentle increase in the computational complexity of the determinant amplitudes. The scaling with $\alpha$ is linear, as the evaluation of the elements hidden sub-matrix requires $\mathcal{O} \big( \tilde{N}(M\cdot \alpha M + \alpha M \cdot (N + \tilde{N}) )\big)$ operations, coming from the two afine transformations of the fully connected neural networks with a single hidden layer. For reference, the scaling of the evaluation of the neural-network backflow from~\cite{Luo2019backflow} is $\mathcal{O}(N^3)$, from the evaluation of the determinant of multi-particle orbitals, while the evaluation of the matrix elements that enter the determinant requires to store $M$ distinct fully connected neural networks and $\mathcal{O}\big(N(M\cdot \alpha M + \alpha M \cdot (N) \big)$ operations. This makes the asymptotic scaling of the hidden fermion determinant state with $N$, $M$ and $\alpha$ identical to the scaling of the  neural-network backflow

In principle, deeper architectures provide a greater expressive power than their shallower counterparts~\cite{montufar2014deepernets}, at the expense of a higher computational cost. We observe that, while deeper architectures provide marginal gains in the energy error, increasing the number of hidden fermions yields a greater impact on the accuracy of the \textit{ansatz} (see the SI~\cite{SI} for a detailed study of the effect of increasing $\tilde{N}$ and the depth of the neural networks in the accuracy of the \textit{ansatz}). 

 Benchmarks on physically motivated constraint functions were also performed (see the SI~\cite{SI} for details). Our experiments reveal that parametrizing $f$ is advantageous compared to the physically inspired rigid rules, which show a marginal improvement in accuracy as compared to the Slater-Jastrow state.

\begin{figure*}
            \includegraphics[width=.95\linewidth]{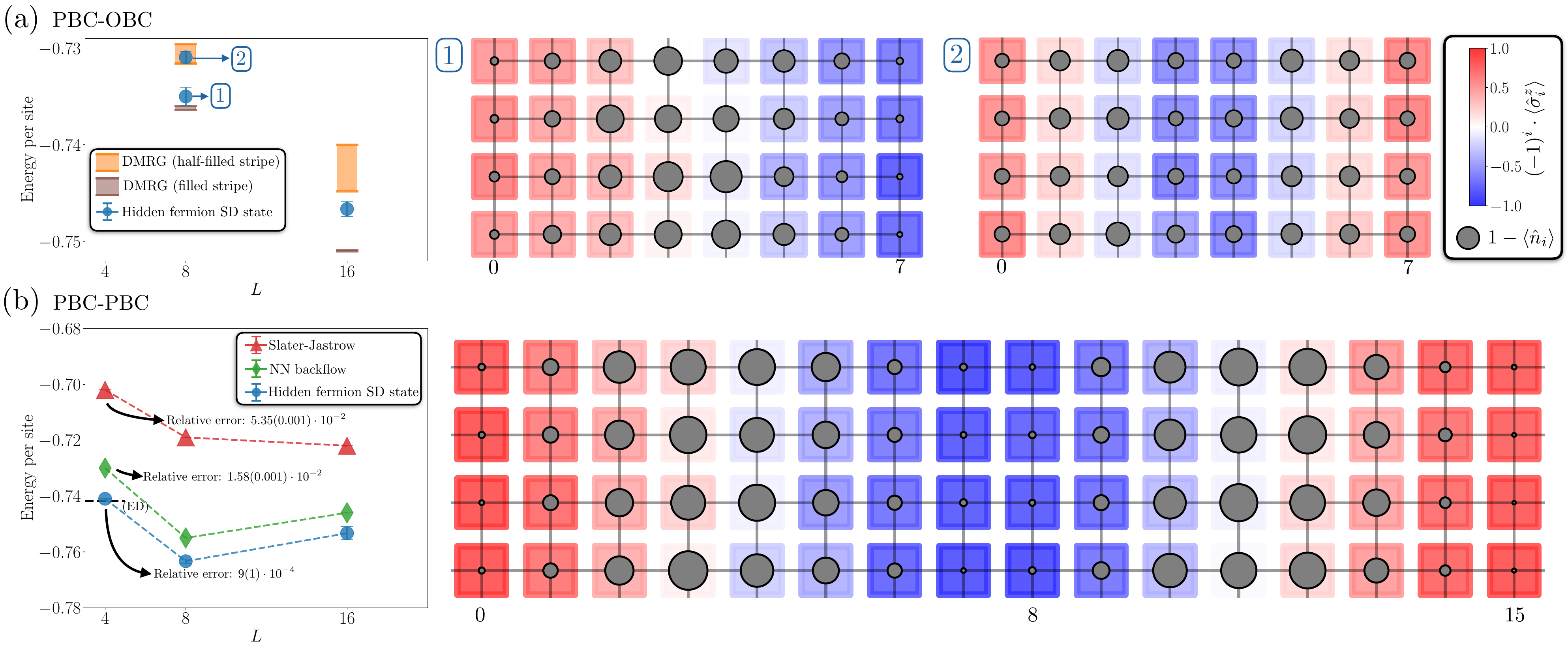}
            \caption{\label{fig_04: cylinders}  Energy per site and competing charge and spin orders in the $4\times L$ rectangular lattice at $1/8$ hole doping ($n = 0.875$) and $U = 8$. \textbf{(a)} Periodic boundary conditions on the short side of the cylinder and open on the long side (PBC-OBC). The left panel compares the hidden fermion determinant state energies with DMRG energies. The width of the DMRG symbols shows the range of converged variational energies for different bond dimensions used in Ref.~\cite{Zheng2017stripes}. For $L = 8$, blue points labelled as [1] and [2] correspond to filled and half filled stripes. The right panel shows the hole and staggered spin distribution for both meta-stable configurations. The diameter of the grey circles is proportional to the hole density. \textbf{(b)} Periodic boundary conditions along both sides of the rectangles (PBC-PBC). The left panel compares the hidden fermion determinant state energies with the Slater-Jastrow and neural network backflow {\it ans\"{a}tze} (from Ref.~\cite{Luo2019backflow}). The dashed horizontal line marks the ED ($4\times 4$ with PBCs from Ref.~\cite{Dagotto1992ED}) energy. In the $4\times 4$ lattice the relative error in the ground state energy is displayed for each {\it ansatz}. The right panel shows the hole and staggered spin distributions in the $4\times 16$ lattice. }
\end{figure*} 

At $n = 5/8$ average site occupation we can compare the relative error in the energy against the \textit{state-of-the-art} {\it ansatz}  from Ref.~\cite{Hui-Hai2017treeTN}. $n = 5/8$, which corresponds to $N = 10$, is the first closed shell for the model under consideration in the non-interacting limit. Fig.~\ref{fig_03: 4X4 benchmark} (b) shows the relative error in the ground-state energy as a function of $U$. The error does not increase monotonically with the value of $U$, as standard wave function {\it ans\"{a}tze}  do. This is shown by the reference errors displayed by the Slater-Jastrow and Slater-RBM {\it ans\"{a}tze}. Remarkably, the single Slater determinant {\it ansatz} with the parametrized constraint function outperforms (by a factor of two in the relative error) the relative error reported in Ref.~\cite{Hui-Hai2017treeTN} that uses the \textit{state-of-the-art} {\it ansatz} consisting on a pairing reference state multiplied by Jastrow, Gutzwiller and doublon-holon correlation factors as well as fat tree tensor network of bond dimension 16, all projected into the zero momentum singlet subspace, with enforced $C_4$ rotational symmetry. Remarkably, while the result from Ref.~\cite{Hui-Hai2017treeTN} relies on the projection of the trial state onto given symmetry sectors, our \textit{ansatz} achieves better accuracy with no symmetry projections. Symmetry projections are an independent avenue to improve the accuracy. So far, we have only considered the increase of the neural-network complexity to obtain better trial states. Not surprisingly, our \textit{ansatz} is further improved when projected to relevant symmetry subspaces after its convergence, as shown in panel (b) of Fig.~\ref{fig_03: 4X4 benchmark}. 


See the SI~\cite{SI} for more benchmarks in square geometries at half  filling, where we compare the variational energies from the hidden fermion determinant state at increasingly larger system sizes with Auxiliary Field Quantum Monte Carlo (AFQMC) calculations~\cite{Qin2026AFQMCBenchmark}. Our energies are in better agreement to AFQMC than those obtained with the neural-network Jastrow wave function from Ref.~\cite{Nomura2017slaterRBM}.

\subsection{Increasing system size and stripe order at $1/8$ hole-doped rectangular geometries}

To conclude this work, we investigate the validity of the proposed wave function {\it ansatz} on increasingly larger system sizes  in rectangular geometries. In particular, we choose rectangular lattices of dimensions $L\times 4$, with $L = \{4, 8, 16\}$.  We focus on the $1/8$ hole doped and zero total magnetization subspace, where, in the strong coupling regime, the ground state is expected to show hole stripes every eight lattice sites across the long side of the rectangle ($\lambda = 8$). The high hole density regions coincide with domain walls in the antiferromagnetic order~\cite{Zheng2017stripes, Wietek2021stripes}. In this section we study the particular case of $U = 8$. 
For this particular choice of coupling constant and filling, previous works have found different competing orders close in energy to the $\lambda = 8$ stripe order~\cite{Zheng2017stripes}. To guide the wave function \textit{ansatz} towards the $\lambda = 8$ stripe order, we add a soft mean-field constraint to the $\phi_{\rm v}$ matrix of orbitals. A $M\times N$ matrix is added to the variational $\phi_{\rm v}$. This matrix has zeros everywhere except for a single entry in every column. These entries are filled up with a constant factor $\mathcal{S}$ that multiplies $\max (|\phi_{\rm v}|)$, following the charge and spin order described above. This forces each of the $N$ visible orbitals to peak in a certain position of the physical lattice. The value of $\mathcal{S}$ is a hyper-parameter. The wave function is optimized with this constraint until its convergence, then the guiding matrix is merged into $\phi_{\rm v}$ as part of the variational parameters and the energy optimization is continued. 

A good trade-off between accuracy and computational resource use is achieved by the addition of $\tilde{N} = 16$ hidden fermions for all system sizes. We also consider a two layer fully connected neural network of hidden-unit density $\alpha = \{60, 14, 6\}$ for the $L = \{4, 8, 16\}$ sizes respectively, to parametrize the rows of the hidden sub-matrix.

We investigate the case with periodic boundary conditions on the short side of the rectangle and open boundary conditions on the long side (PBC-OBC). The left panel of Fig.~\ref{fig_04: cylinders} (a) shows the energy per site as a function of $L$ and the comparison with DMRG variational energies used in Ref.~\cite{Zheng2017stripes}. The DMRG algorithm finds two meta-stable solutions, one with half-filled stripes and one with filled stripes. In the $L = 8$ case, we can stabilize both meta-stable arrangements by tuning the value of $\mathcal{S}$. $\mathcal{S} = 0$ or small values of $\mathcal{S}$ lead to a half-filled stripe configuration of higher energy. In this system size the charge distribution shows high hole density every four sites, coinciding with domain walls in the antiferromagnetic order. Larger values of $\mathcal{S}$ yield a filled stripe configuration, showing only one stripe of high hole density in the system, that coincides with a domain wall in the antiferromagnetic order. The charge and spin configurations for the two competing orders are shown on the right panel of Fig.~\ref{fig_04: cylinders} (a). They are in good agreement with the DMRG hole distributions. The variational energy from the hidden fermion determinant state is in good agreement with the DMRG energies. At $L = 16$, the best variational energy found by tuning $\mathcal{S}$ lies between the DMRG energies that correspond to the half and filled stripes. 
Our method, not being specifically tailored to quasi one-dimensional problems, does not outperform DMRG in this particular lattice geometry.

A more interesting case is the addition of periodic boundary conditions along the long side of the rectangle, a situation that is not amenable to DMRG calculations due to its computational cost. The left panel of Fig.~\ref{fig_04: cylinders} (b) shows the energy per site as a function of $L$ and the comparison with the ED energy in the $4\times 4$ system. The hidden fermion determinant with a parametrized constraint function achieves a significantly lower energy than both the standard Slater-Jastrow {\it ansatz} and the state-of-the-art neural network backflow {\it ansatz} from Ref.~\cite{Luo2019backflow}. In the $4\times 4$ lattice, the relative error in the energy is reduced by over one order of magnitude compared to the neural network backflow wave function. In the $4\times 8$ and $4 \times 16$ lattices the energy per site is noticeably lower than it from the neural network backflow and Slater-Jastrow {\it ans\"{a}tze}. These results demonstrate the escalability of the proposed formalism, which outperforms existing state-of-the-art wave function {\it ans\"{a}tze}  used in the field. For these cases we set $\mathcal{S} = 3$.

In addition, we analyze the hole density and staggered spin density distributions in the largest system size ($4\times 16$) in the right panel of  Fig.~\ref{fig_04: cylinders} (b). The hole density distribution shows repeating maxima separated by $8$ lattice sites. Coinciding with the maxima in the hole density, the antiferromagnetic order displays a domain wall. The amplitude of the staggered magnetization is modulated along the long side of the rectangles. These features are consistent with observations from previous studies~\cite{Zheng2017stripes, Wietek2021stripes} coming from different many-body numerical methods, further validating the accuracy of the hidden fermion determinant state to find good approximations to the highly correlated ground states of complex Hamiltonians.

\section{Conclusions}

In this paper we have shown that the variational treatment of interacting electrons in an augmented Fock space can be highly beneficial to improve the generality of the wave function {\it ansatz}, especially in the strong correlation limit. 
We found that key elements for the success of this approach are to optimize the constraint function relating the 
enlarged and physical Hilbert spaces, as well as treating this constraint exactly. 
A simple Slater determinant state in the augmented Fock space is found to  
provide an extremely expressive wave function {\it ansatz}, which we proved to be universal. 
The optimization of the constraint function and of the hidden sector orbital amplitudes was 
performed using a neural network representation. 
We presented numerical experiments which show that the proposed wave function is competitive with the  \textit{state-of-the-art} variational accuracy in the ground state of the Hubbard model in the square lattice. 
Furthermore, in contrast to standard variational approaches~\cite{Nomura2017slaterRBM, Hui-Hai2017treeTN}, the accuracy of this {\it ansatz} does not rely on imposing symmetries, 
potentially allowing for a great level of accuracy on systems with a small number of symmetries. This also opens the possibility of applying our approach to systems without an underlying lattice, with potential applications to quantum chemistry, nuclear physics and materials science.

In particular, we envision the accurate calculation of the ground-state properties of molecular Hamiltonians, which in the Molecular Orbital basis lacks both an underlying lattice and exploitable symmetries. The connection between a compact representation of a CI wave function with all possible single-, double-, up to  $\tilde{N}$-tuple excitations provides an accurate post Hatree-Fock starting point for the trial state. 
The formalism introduced in this manuscript is also well-suited for models with quenched disorder, 
such as 
models of interacting fermions on fully connected lattices 
with random exchange interactions~\cite{SYK_review}. 
While exact solutions of such models are available in the large-$M$ limit when the spin symmetry  
is extended to $SU(M)$, the physical $SU(2)$ case requires computational approaches. 
Similarly to Molecular Hamiltonians, this class of Hamiltonians also lacks any translational symmetries 
that can be exploited to improve the accuracy of traditional wave-function {\it ans\"{a}tze}.

\acknowledgements
The Flatiron Institute is a division of the Simons Foundation. GC is supported by the Swiss National Science Foundation under Grant No. 200021\_200336.  The authors acknowledge Steven White for providing DMRG raw data of the energies of Ref.~\cite{Zheng2017stripes} of cylindrical geometries.

\bibliography{references.bib}

\begin{thebibliography}{53}%
\makeatletter
\providecommand \@ifxundefined [1]{%
 \@ifx{#1\undefined}
}%
\providecommand \@ifnum [1]{%
 \ifnum #1\expandafter \@firstoftwo
 \else \expandafter \@secondoftwo
 \fi
}%
\providecommand \@ifx [1]{%
 \ifx #1\expandafter \@firstoftwo
 \else \expandafter \@secondoftwo
 \fi
}%
\providecommand \natexlab [1]{#1}%
\providecommand \enquote  [1]{``#1''}%
\providecommand \bibnamefont  [1]{#1}%
\providecommand \bibfnamefont [1]{#1}%
\providecommand \citenamefont [1]{#1}%
\providecommand \href@noop [0]{\@secondoftwo}%
\providecommand \href [0]{\begingroup \@sanitize@url \@href}%
\providecommand \@href[1]{\@@startlink{#1}\@@href}%
\providecommand \@@href[1]{\endgroup#1\@@endlink}%
\providecommand \@sanitize@url [0]{\catcode `\\12\catcode `\$12\catcode
  `\&12\catcode `\#12\catcode `\^12\catcode `\_12\catcode `\%12\relax}%
\providecommand \@@startlink[1]{}%
\providecommand \@@endlink[0]{}%
\providecommand \url  [0]{\begingroup\@sanitize@url \@url }%
\providecommand \@url [1]{\endgroup\@href {#1}{\urlprefix }}%
\providecommand \urlprefix  [0]{URL }%
\providecommand \Eprint [0]{\href }%
\providecommand \doibase [0]{http://dx.doi.org/}%
\providecommand \selectlanguage [0]{\@gobble}%
\providecommand \bibinfo  [0]{\@secondoftwo}%
\providecommand \bibfield  [0]{\@secondoftwo}%
\providecommand \translation [1]{[#1]}%
\providecommand \BibitemOpen [0]{}%
\providecommand \bibitemStop [0]{}%
\providecommand \bibitemNoStop [0]{.\EOS\space}%
\providecommand \EOS [0]{\spacefactor3000\relax}%
\providecommand \BibitemShut  [1]{\csname bibitem#1\endcsname}%
\let\auto@bib@innerbib\@empty
\bibitem [{not()}]{note_slave}%
  \BibitemOpen
  \href@noop {} {}\bibinfo {note} {In the condensed matter physics literature,
  the term `slave-particle' representations has been used historically to
  denote approaches in which the physical Hilbert space is viewed as the
  projection of an enlarged Hilbert space, often in conjunction with a
  subsequent mean-field treatment. In this work, we deem appropriate to use a
  different terminology and denote by `visible' and `hidden' the previously
  called `auxiliary' and `slave' degrees of freedom, respectively.}\BibitemShut
  {Stop}%
\bibitem [{\citenamefont {Lieb}(1973)}]{lieb1973classical}%
  \BibitemOpen
  \bibfield  {author} {\bibinfo {author} {\bibfnamefont {Elliott~H}\
  \bibnamefont {Lieb}},\ }\bibfield  {title} {\enquote {\bibinfo {title} {The
  classical limit of quantum spin systems},}\ }\href@noop {} {\bibfield
  {journal} {\bibinfo  {journal} {Communications in Mathematical Physics}\
  }\textbf {\bibinfo {volume} {31}},\ \bibinfo {pages} {327--340} (\bibinfo
  {year} {1973})}\BibitemShut {NoStop}%
\bibitem [{\citenamefont {Bravyi}\ \emph {et~al.}(2019)\citenamefont {Bravyi},
  \citenamefont {Gosset}, \citenamefont {K{\"o}nig},\ and\ \citenamefont
  {Temme}}]{bravyi2019approximation}%
  \BibitemOpen
  \bibfield  {author} {\bibinfo {author} {\bibfnamefont {Sergey}\ \bibnamefont
  {Bravyi}}, \bibinfo {author} {\bibfnamefont {David}\ \bibnamefont {Gosset}},
  \bibinfo {author} {\bibfnamefont {Robert}\ \bibnamefont {K{\"o}nig}}, \ and\
  \bibinfo {author} {\bibfnamefont {Kristan}\ \bibnamefont {Temme}},\
  }\bibfield  {title} {\enquote {\bibinfo {title} {Approximation algorithms for
  quantum many-body problems},}\ }\href@noop {} {\bibfield  {journal} {\bibinfo
   {journal} {Journal of Mathematical Physics}\ }\textbf {\bibinfo {volume}
  {60}},\ \bibinfo {pages} {032203} (\bibinfo {year} {2019})}\BibitemShut
  {NoStop}%
\bibitem [{\citenamefont {Hyatt}\ and\ \citenamefont
  {Stoudenmire}(2020)}]{hyatt20202D_tensor_nets}%
  \BibitemOpen
  \bibfield  {author} {\bibinfo {author} {\bibfnamefont {Katharine}\
  \bibnamefont {Hyatt}}\ and\ \bibinfo {author} {\bibfnamefont {E.~M.}\
  \bibnamefont {Stoudenmire}},\ }\href@noop {} {\enquote {\bibinfo {title}
  {Dmrg approach to optimizing two-dimensional tensor networks},}\ } (\bibinfo
  {year} {2020}),\ \Eprint {http://arxiv.org/abs/1908.08833} {arXiv:1908.08833
  [cond-mat.str-el]} \BibitemShut {NoStop}%
\bibitem [{\citenamefont {Choo}\ \emph {et~al.}(2019)\citenamefont {Choo},
  \citenamefont {Neupert},\ and\ \citenamefont {Carleo}}]{Choo2019J1J2}%
  \BibitemOpen
  \bibfield  {author} {\bibinfo {author} {\bibfnamefont {Kenny}\ \bibnamefont
  {Choo}}, \bibinfo {author} {\bibfnamefont {Titus}\ \bibnamefont {Neupert}}, \
  and\ \bibinfo {author} {\bibfnamefont {Giuseppe}\ \bibnamefont {Carleo}},\
  }\bibfield  {title} {\enquote {\bibinfo {title} {Two-dimensional frustrated
  ${J}_{1}\text{\ensuremath{-}}{J}_{2}$ model studied with neural network
  quantum states},}\ }\href {\doibase 10.1103/PhysRevB.100.125124} {\bibfield
  {journal} {\bibinfo  {journal} {Phys. Rev. B}\ }\textbf {\bibinfo {volume}
  {100}},\ \bibinfo {pages} {125124} (\bibinfo {year} {2019})}\BibitemShut
  {NoStop}%
\bibitem [{\citenamefont {Choo}\ \emph {et~al.}(2020)\citenamefont {Choo},
  \citenamefont {Mezzacapo},\ and\ \citenamefont {Carleo}}]{Choo2020chemistry}%
  \BibitemOpen
  \bibfield  {author} {\bibinfo {author} {\bibfnamefont {Kenny}\ \bibnamefont
  {Choo}}, \bibinfo {author} {\bibfnamefont {Antonio}\ \bibnamefont
  {Mezzacapo}}, \ and\ \bibinfo {author} {\bibfnamefont {Giuseppe}\
  \bibnamefont {Carleo}},\ }\bibfield  {title} {\enquote {\bibinfo {title}
  {Fermionic neural-network states for ab-initio electronic structure},}\
  }\href {\doibase 10.1038/s41467-020-15724-9} {\bibfield  {journal} {\bibinfo
  {journal} {Nature Communications}\ }\textbf {\bibinfo {volume} {11}},\
  \bibinfo {pages} {2368} (\bibinfo {year} {2020})}\BibitemShut {NoStop}%
\bibitem [{\citenamefont {Yoshioka}\ \emph {et~al.}(2021)\citenamefont
  {Yoshioka}, \citenamefont {Mizukami},\ and\ \citenamefont
  {Nori}}]{Yoshioka2021solids}%
  \BibitemOpen
  \bibfield  {author} {\bibinfo {author} {\bibfnamefont {Nobuyuki}\
  \bibnamefont {Yoshioka}}, \bibinfo {author} {\bibfnamefont {Wataru}\
  \bibnamefont {Mizukami}}, \ and\ \bibinfo {author} {\bibfnamefont {Franco}\
  \bibnamefont {Nori}},\ }\bibfield  {title} {\enquote {\bibinfo {title}
  {Solving quasiparticle band spectra of real solids using neural-network
  quantum states},}\ }\href {\doibase 10.1038/s42005-021-00609-0} {\bibfield
  {journal} {\bibinfo  {journal} {Communications Physics}\ }\textbf {\bibinfo
  {volume} {4}},\ \bibinfo {pages} {106} (\bibinfo {year} {2021})}\BibitemShut
  {NoStop}%
\bibitem [{\citenamefont {Bennewitz}\ \emph {et~al.}(2021)\citenamefont
  {Bennewitz}, \citenamefont {Hopfmueller}, \citenamefont {Kulchytskyy},
  \citenamefont {Carrasquilla},\ and\ \citenamefont
  {Ronagh}}]{bennewitz2021VMC+VQE}%
  \BibitemOpen
  \bibfield  {author} {\bibinfo {author} {\bibfnamefont {Elizabeth~R.}\
  \bibnamefont {Bennewitz}}, \bibinfo {author} {\bibfnamefont {Florian}\
  \bibnamefont {Hopfmueller}}, \bibinfo {author} {\bibfnamefont {Bohdan}\
  \bibnamefont {Kulchytskyy}}, \bibinfo {author} {\bibfnamefont {Juan}\
  \bibnamefont {Carrasquilla}}, \ and\ \bibinfo {author} {\bibfnamefont
  {Pooya}\ \bibnamefont {Ronagh}},\ }\href@noop {} {\enquote {\bibinfo {title}
  {Neural error mitigation of near-term quantum simulations},}\ } (\bibinfo
  {year} {2021}),\ \Eprint {http://arxiv.org/abs/2105.08086} {arXiv:2105.08086
  [quant-ph]} \BibitemShut {NoStop}%
\bibitem [{\citenamefont {Nomura}\ \emph {et~al.}(2017)\citenamefont {Nomura},
  \citenamefont {Darmawan}, \citenamefont {Yamaji},\ and\ \citenamefont
  {Imada}}]{Nomura2017slaterRBM}%
  \BibitemOpen
  \bibfield  {author} {\bibinfo {author} {\bibfnamefont {Yusuke}\ \bibnamefont
  {Nomura}}, \bibinfo {author} {\bibfnamefont {Andrew~S.}\ \bibnamefont
  {Darmawan}}, \bibinfo {author} {\bibfnamefont {Youhei}\ \bibnamefont
  {Yamaji}}, \ and\ \bibinfo {author} {\bibfnamefont {Masatoshi}\ \bibnamefont
  {Imada}},\ }\bibfield  {title} {\enquote {\bibinfo {title} {Restricted
  boltzmann machine learning for solving strongly correlated quantum
  systems},}\ }\href {\doibase 10.1103/PhysRevB.96.205152} {\bibfield
  {journal} {\bibinfo  {journal} {Phys. Rev. B}\ }\textbf {\bibinfo {volume}
  {96}},\ \bibinfo {pages} {205152} (\bibinfo {year} {2017})}\BibitemShut
  {NoStop}%
\bibitem [{\citenamefont {Luo}\ and\ \citenamefont
  {Clark}(2019)}]{Luo2019backflow}%
  \BibitemOpen
  \bibfield  {author} {\bibinfo {author} {\bibfnamefont {Di}~\bibnamefont
  {Luo}}\ and\ \bibinfo {author} {\bibfnamefont {Bryan~K.}\ \bibnamefont
  {Clark}},\ }\bibfield  {title} {\enquote {\bibinfo {title} {Backflow
  transformations via neural networks for quantum many-body wave functions},}\
  }\href {\doibase 10.1103/PhysRevLett.122.226401} {\bibfield  {journal}
  {\bibinfo  {journal} {Phys. Rev. Lett.}\ }\textbf {\bibinfo {volume} {122}},\
  \bibinfo {pages} {226401} (\bibinfo {year} {2019})}\BibitemShut {NoStop}%
\bibitem [{\citenamefont {Stokes}\ \emph {et~al.}(2020)\citenamefont {Stokes},
  \citenamefont {Moreno}, \citenamefont {Pnevmatikakis},\ and\ \citenamefont
  {Carleo}}]{Stokes2020}%
  \BibitemOpen
  \bibfield  {author} {\bibinfo {author} {\bibfnamefont {James}\ \bibnamefont
  {Stokes}}, \bibinfo {author} {\bibfnamefont {Javier~Robledo}\ \bibnamefont
  {Moreno}}, \bibinfo {author} {\bibfnamefont {Eftychios~A.}\ \bibnamefont
  {Pnevmatikakis}}, \ and\ \bibinfo {author} {\bibfnamefont {Giuseppe}\
  \bibnamefont {Carleo}},\ }\bibfield  {title} {\enquote {\bibinfo {title}
  {Phases of two-dimensional spinless lattice fermions with first-quantized
  deep neural-network quantum states},}\ }\href {\doibase
  10.1103/PhysRevB.102.205122} {\bibfield  {journal} {\bibinfo  {journal}
  {Phys. Rev. B}\ }\textbf {\bibinfo {volume} {102}},\ \bibinfo {pages}
  {205122} (\bibinfo {year} {2020})}\BibitemShut {NoStop}%
\bibitem [{\citenamefont {Pfau}\ \emph {et~al.}(2020)\citenamefont {Pfau},
  \citenamefont {Spencer}, \citenamefont {Matthews},\ and\ \citenamefont
  {Foulkes}}]{Pfau2020Ferminet}%
  \BibitemOpen
  \bibfield  {author} {\bibinfo {author} {\bibfnamefont {David}\ \bibnamefont
  {Pfau}}, \bibinfo {author} {\bibfnamefont {James~S.}\ \bibnamefont
  {Spencer}}, \bibinfo {author} {\bibfnamefont {Alexander G. D.~G.}\
  \bibnamefont {Matthews}}, \ and\ \bibinfo {author} {\bibfnamefont {W.~M.~C.}\
  \bibnamefont {Foulkes}},\ }\bibfield  {title} {\enquote {\bibinfo {title} {Ab
  initio solution of the many-electron schr\"odinger equation with deep neural
  networks},}\ }\href {\doibase 10.1103/PhysRevResearch.2.033429} {\bibfield
  {journal} {\bibinfo  {journal} {Phys. Rev. Research}\ }\textbf {\bibinfo
  {volume} {2}},\ \bibinfo {pages} {033429} (\bibinfo {year}
  {2020})}\BibitemShut {NoStop}%
\bibitem [{\citenamefont {Spencer}\ \emph {et~al.}(2020)\citenamefont
  {Spencer}, \citenamefont {Pfau}, \citenamefont {Botev},\ and\ \citenamefont
  {Foulkes}}]{Spencer2020better}%
  \BibitemOpen
  \bibfield  {author} {\bibinfo {author} {\bibfnamefont {James~S.}\
  \bibnamefont {Spencer}}, \bibinfo {author} {\bibfnamefont {David}\
  \bibnamefont {Pfau}}, \bibinfo {author} {\bibfnamefont {Aleksandar}\
  \bibnamefont {Botev}}, \ and\ \bibinfo {author} {\bibfnamefont {W.~M.~C.}\
  \bibnamefont {Foulkes}},\ }\bibfield  {title} {\enquote {\bibinfo {title}
  {Better, faster fermionic neural networks},}\ }\href@noop {} {\  (\bibinfo
  {year} {2020})},\ \Eprint {http://arxiv.org/abs/2011.07125} {arXiv:2011.07125
  [physics.comp-ph]} \BibitemShut {NoStop}%
\bibitem [{\citenamefont {Hermann}\ \emph {et~al.}(2020)\citenamefont
  {Hermann}, \citenamefont {Sch{\"a}tzle},\ and\ \citenamefont
  {No{\'e}}}]{Hermann2020Paulinet}%
  \BibitemOpen
  \bibfield  {author} {\bibinfo {author} {\bibfnamefont {Jan}\ \bibnamefont
  {Hermann}}, \bibinfo {author} {\bibfnamefont {Zeno}\ \bibnamefont
  {Sch{\"a}tzle}}, \ and\ \bibinfo {author} {\bibfnamefont {Frank}\
  \bibnamefont {No{\'e}}},\ }\bibfield  {title} {\enquote {\bibinfo {title}
  {Deep-neural-network solution of the electronic schr{\"o}dinger equation},}\
  }\href {\doibase 10.1038/s41557-020-0544-y} {\bibfield  {journal} {\bibinfo
  {journal} {Nature Chemistry}\ }\textbf {\bibinfo {volume} {12}},\ \bibinfo
  {pages} {891--897} (\bibinfo {year} {2020})}\BibitemShut {NoStop}%
\bibitem [{\citenamefont {Inui}\ \emph {et~al.}(2021)\citenamefont {Inui},
  \citenamefont {Kato},\ and\ \citenamefont
  {Motome}}]{Inui2021determinantfree}%
  \BibitemOpen
  \bibfield  {author} {\bibinfo {author} {\bibfnamefont {Koji}\ \bibnamefont
  {Inui}}, \bibinfo {author} {\bibfnamefont {Yasuyuki}\ \bibnamefont {Kato}}, \
  and\ \bibinfo {author} {\bibfnamefont {Yukitoshi}\ \bibnamefont {Motome}},\
  }\href@noop {} {\enquote {\bibinfo {title} {Determinant-free fermionic wave
  function using feed-forward neural networks},}\ } (\bibinfo {year} {2021}),\
  \Eprint {http://arxiv.org/abs/2108.08631} {arXiv:2108.08631
  [cond-mat.str-el]} \BibitemShut {NoStop}%
\bibitem [{\citenamefont {Barnes}(1976)}]{barnes_1976}%
  \BibitemOpen
  \bibfield  {author} {\bibinfo {author} {\bibfnamefont {S~E}\ \bibnamefont
  {Barnes}},\ }\bibfield  {title} {\enquote {\bibinfo {title} {New method for
  the anderson model},}\ }\href {\doibase 10.1088/0305-4608/6/7/018} {\ \textbf
  {\bibinfo {volume} {6}},\ \bibinfo {pages} {1375--1383} (\bibinfo {year}
  {1976})}\BibitemShut {NoStop}%
\bibitem [{\citenamefont {Barnes}(1977)}]{barnes_1977}%
  \BibitemOpen
  \bibfield  {author} {\bibinfo {author} {\bibfnamefont {S~E}\ \bibnamefont
  {Barnes}},\ }\bibfield  {title} {\enquote {\bibinfo {title} {New method for
  the anderson model. {II}. the u=0 limit},}\ }\href {\doibase
  10.1088/0305-4608/7/12/022} {\ \textbf {\bibinfo {volume} {7}},\ \bibinfo
  {pages} {2637--2647} (\bibinfo {year} {1977})}\BibitemShut {NoStop}%
\bibitem [{\citenamefont {Coleman}(1984)}]{coleman_1984}%
  \BibitemOpen
  \bibfield  {author} {\bibinfo {author} {\bibfnamefont {Piers}\ \bibnamefont
  {Coleman}},\ }\bibfield  {title} {\enquote {\bibinfo {title} {New approach to
  the mixed-valence problem},}\ }\href {\doibase 10.1103/PhysRevB.29.3035}
  {\bibfield  {journal} {\bibinfo  {journal} {Phys. Rev. B}\ }\textbf {\bibinfo
  {volume} {29}},\ \bibinfo {pages} {3035--3044} (\bibinfo {year}
  {1984})}\BibitemShut {NoStop}%
\bibitem [{\citenamefont {Kotliar}\ and\ \citenamefont
  {Ruckenstein}(1986)}]{Kotliar1986SlaveBosons}%
  \BibitemOpen
  \bibfield  {author} {\bibinfo {author} {\bibfnamefont {Gabriel}\ \bibnamefont
  {Kotliar}}\ and\ \bibinfo {author} {\bibfnamefont {Andrei~E.}\ \bibnamefont
  {Ruckenstein}},\ }\bibfield  {title} {\enquote {\bibinfo {title} {New
  functional integral approach to strongly correlated fermi systems: The
  gutzwiller approximation as a saddle point},}\ }\href {\doibase
  10.1103/PhysRevLett.57.1362} {\bibfield  {journal} {\bibinfo  {journal}
  {Phys. Rev. Lett.}\ }\textbf {\bibinfo {volume} {57}},\ \bibinfo {pages}
  {1362--1365} (\bibinfo {year} {1986})}\BibitemShut {NoStop}%
\bibitem [{\citenamefont {Kotliar}(1995)}]{kotliar_largeN}%
  \BibitemOpen
  \bibfield  {author} {\bibinfo {author} {\bibfnamefont {G.}~\bibnamefont
  {Kotliar}},\ }\bibfield  {title} {\enquote {\bibinfo {title} {The large n
  expansion and the strong correlation problem},}\ }in\ \href@noop {} {\emph
  {\bibinfo {booktitle} {Strongly Interacting Fermions and High-T$_c$
  superconductivity}}},\ \bibinfo {series and number} {Les Houches, Session
  LVI},\ \bibinfo {editor} {edited by\ \bibinfo {editor} {\bibfnamefont
  {B.}~\bibnamefont {Doucot}}\ and\ \bibinfo {editor} {\bibfnamefont
  {J.}~\bibnamefont {Zinn-Justin}}}\ (\bibinfo  {publisher} {Elsevier},\
  \bibinfo {year} {1995})\ p.\ \bibinfo {pages} {197}\BibitemShut {NoStop}%
\bibitem [{\citenamefont {Li}\ \emph {et~al.}(1989)\citenamefont {Li},
  \citenamefont {W\"olfle},\ and\ \citenamefont {Hirschfeld}}]{li_rotinv_1989}%
  \BibitemOpen
  \bibfield  {author} {\bibinfo {author} {\bibfnamefont {T.}~\bibnamefont
  {Li}}, \bibinfo {author} {\bibfnamefont {P.}~\bibnamefont {W\"olfle}}, \ and\
  \bibinfo {author} {\bibfnamefont {P.~J.}\ \bibnamefont {Hirschfeld}},\
  }\bibfield  {title} {\enquote {\bibinfo {title} {Spin-rotation-invariant
  slave-boson approach to the hubbard model},}\ }\href {\doibase
  10.1103/PhysRevB.40.6817} {\bibfield  {journal} {\bibinfo  {journal} {Phys.
  Rev. B}\ }\textbf {\bibinfo {volume} {40}},\ \bibinfo {pages} {6817--6821}
  (\bibinfo {year} {1989})}\BibitemShut {NoStop}%
\bibitem [{\citenamefont {Frésard}\ and\ \citenamefont
  {Wölfle}(1992)}]{fresard_1992}%
  \BibitemOpen
  \bibfield  {author} {\bibinfo {author} {\bibfnamefont {R.}~\bibnamefont
  {Frésard}}\ and\ \bibinfo {author} {\bibfnamefont {P.}~\bibnamefont
  {Wölfle}},\ }\bibfield  {title} {\enquote {\bibinfo {title} {Unified slave
  boson representation of spin and charge degrees of freedom for strongly
  correlated fermi systems},}\ }\href {\doibase 10.1142/S0217979292000414}
  {\bibfield  {journal} {\bibinfo  {journal} {International Journal of Modern
  Physics B}\ }\textbf {\bibinfo {volume} {06}},\ \bibinfo {pages} {685--704}
  (\bibinfo {year} {1992})}\BibitemShut {NoStop}%
\bibitem [{\citenamefont {Jayaprakash}\ \emph {et~al.}(1989)\citenamefont
  {Jayaprakash}, \citenamefont {Krishnamurthy},\ and\ \citenamefont
  {Sarker}}]{slavefermion1}%
  \BibitemOpen
  \bibfield  {author} {\bibinfo {author} {\bibfnamefont {C.}~\bibnamefont
  {Jayaprakash}}, \bibinfo {author} {\bibfnamefont {H.~R.}\ \bibnamefont
  {Krishnamurthy}}, \ and\ \bibinfo {author} {\bibfnamefont {Sanjoy}\
  \bibnamefont {Sarker}},\ }\bibfield  {title} {\enquote {\bibinfo {title}
  {Mean-field theory for the t-j model},}\ }\href {\doibase
  10.1103/PhysRevB.40.2610} {\bibfield  {journal} {\bibinfo  {journal} {Phys.
  Rev. B}\ }\textbf {\bibinfo {volume} {40}},\ \bibinfo {pages} {2610--2613}
  (\bibinfo {year} {1989})}\BibitemShut {NoStop}%
\bibitem [{\citenamefont {Kane}\ \emph {et~al.}(1990)\citenamefont {Kane},
  \citenamefont {Lee}, \citenamefont {Ng}, \citenamefont {Chakraborty},\ and\
  \citenamefont {Read}}]{slavefermion2}%
  \BibitemOpen
  \bibfield  {author} {\bibinfo {author} {\bibfnamefont {C.~L.}\ \bibnamefont
  {Kane}}, \bibinfo {author} {\bibfnamefont {P.~A.}\ \bibnamefont {Lee}},
  \bibinfo {author} {\bibfnamefont {T.~K.}\ \bibnamefont {Ng}}, \bibinfo
  {author} {\bibfnamefont {B.}~\bibnamefont {Chakraborty}}, \ and\ \bibinfo
  {author} {\bibfnamefont {N.}~\bibnamefont {Read}},\ }\bibfield  {title}
  {\enquote {\bibinfo {title} {Mean-field theory of the spiral phases of a
  doped antiferromagnet},}\ }\href {\doibase 10.1103/PhysRevB.41.2653}
  {\bibfield  {journal} {\bibinfo  {journal} {Phys. Rev. B}\ }\textbf {\bibinfo
  {volume} {41}},\ \bibinfo {pages} {2653--2656} (\bibinfo {year}
  {1990})}\BibitemShut {NoStop}%
\bibitem [{\citenamefont {Florens}\ and\ \citenamefont
  {Georges}(2002)}]{Florens2002SlaveRotor}%
  \BibitemOpen
  \bibfield  {author} {\bibinfo {author} {\bibfnamefont {Serge}\ \bibnamefont
  {Florens}}\ and\ \bibinfo {author} {\bibfnamefont {Antoine}\ \bibnamefont
  {Georges}},\ }\bibfield  {title} {\enquote {\bibinfo {title} {Quantum
  impurity solvers using a slave rotor representation},}\ }\href {\doibase
  10.1103/PhysRevB.66.165111} {\bibfield  {journal} {\bibinfo  {journal} {Phys.
  Rev. B}\ }\textbf {\bibinfo {volume} {66}},\ \bibinfo {pages} {165111}
  (\bibinfo {year} {2002})}\BibitemShut {NoStop}%
\bibitem [{\citenamefont {Florens}\ and\ \citenamefont
  {Georges}(2004)}]{Florens2004Slaverotor}%
  \BibitemOpen
  \bibfield  {author} {\bibinfo {author} {\bibfnamefont {Serge}\ \bibnamefont
  {Florens}}\ and\ \bibinfo {author} {\bibfnamefont {Antoine}\ \bibnamefont
  {Georges}},\ }\bibfield  {title} {\enquote {\bibinfo {title} {Slave-rotor
  mean-field theories of strongly correlated systems and the mott transition in
  finite dimensions},}\ }\href {\doibase 10.1103/PhysRevB.70.035114} {\bibfield
   {journal} {\bibinfo  {journal} {Phys. Rev. B}\ }\textbf {\bibinfo {volume}
  {70}},\ \bibinfo {pages} {035114} (\bibinfo {year} {2004})}\BibitemShut
  {NoStop}%
\bibitem [{\citenamefont {de'Medici}\ \emph {et~al.}(2005)\citenamefont
  {de'Medici}, \citenamefont {Georges},\ and\ \citenamefont
  {Biermann}}]{Medici2005slavespins}%
  \BibitemOpen
  \bibfield  {author} {\bibinfo {author} {\bibfnamefont {L.}~\bibnamefont
  {de'Medici}}, \bibinfo {author} {\bibfnamefont {A.}~\bibnamefont {Georges}},
  \ and\ \bibinfo {author} {\bibfnamefont {S.}~\bibnamefont {Biermann}},\
  }\bibfield  {title} {\enquote {\bibinfo {title} {Orbital-selective mott
  transition in multiband systems: Slave-spin representation and dynamical
  mean-field theory},}\ }\href {\doibase 10.1103/PhysRevB.72.205124} {\bibfield
   {journal} {\bibinfo  {journal} {Phys. Rev. B}\ }\textbf {\bibinfo {volume}
  {72}},\ \bibinfo {pages} {205124} (\bibinfo {year} {2005})}\BibitemShut
  {NoStop}%
\bibitem [{\citenamefont {Lechermann}\ \emph {et~al.}(2007)\citenamefont
  {Lechermann}, \citenamefont {Georges}, \citenamefont {Kotliar},\ and\
  \citenamefont {Parcollet}}]{lechermann_2007}%
  \BibitemOpen
  \bibfield  {author} {\bibinfo {author} {\bibfnamefont {Frank}\ \bibnamefont
  {Lechermann}}, \bibinfo {author} {\bibfnamefont {Antoine}\ \bibnamefont
  {Georges}}, \bibinfo {author} {\bibfnamefont {Gabriel}\ \bibnamefont
  {Kotliar}}, \ and\ \bibinfo {author} {\bibfnamefont {Olivier}\ \bibnamefont
  {Parcollet}},\ }\bibfield  {title} {\enquote {\bibinfo {title} {Rotationally
  invariant slave-boson formalism and momentum dependence of the quasiparticle
  weight},}\ }\href {\doibase 10.1103/PhysRevB.76.155102} {\bibfield  {journal}
  {\bibinfo  {journal} {Phys. Rev. B}\ }\textbf {\bibinfo {volume} {76}},\
  \bibinfo {pages} {155102} (\bibinfo {year} {2007})}\BibitemShut {NoStop}%
\bibitem [{\citenamefont {Lanat\`a}\ \emph {et~al.}(2017)\citenamefont
  {Lanat\`a}, \citenamefont {Lee}, \citenamefont {Yao},\ and\ \citenamefont
  {Dobrosavljevi\ifmmode~\acute{c}\else \'{c}\fi{}}}]{lanata_ghostGA_2017}%
  \BibitemOpen
  \bibfield  {author} {\bibinfo {author} {\bibfnamefont {Nicola}\ \bibnamefont
  {Lanat\`a}}, \bibinfo {author} {\bibfnamefont {Tsung-Han}\ \bibnamefont
  {Lee}}, \bibinfo {author} {\bibfnamefont {Yong-Xin}\ \bibnamefont {Yao}}, \
  and\ \bibinfo {author} {\bibfnamefont {Vladimir}\ \bibnamefont
  {Dobrosavljevi\ifmmode~\acute{c}\else \'{c}\fi{}}},\ }\bibfield  {title}
  {\enquote {\bibinfo {title} {Emergent bloch excitations in mott matter},}\
  }\href {\doibase 10.1103/PhysRevB.96.195126} {\bibfield  {journal} {\bibinfo
  {journal} {Phys. Rev. B}\ }\textbf {\bibinfo {volume} {96}},\ \bibinfo
  {pages} {195126} (\bibinfo {year} {2017})}\BibitemShut {NoStop}%
\bibitem [{\citenamefont {Frank}\ \emph {et~al.}(2021)\citenamefont {Frank},
  \citenamefont {Lee}, \citenamefont {Bhattacharyya}, \citenamefont {Tsang},
  \citenamefont {Quito}, \citenamefont {Dobrosavljevi\ifmmode~\acute{c}\else
  \'{c}\fi{}}, \citenamefont {Christiansen},\ and\ \citenamefont
  {Lanat\`a}}]{frank_ghost_2021}%
  \BibitemOpen
  \bibfield  {author} {\bibinfo {author} {\bibfnamefont {Marius~S.}\
  \bibnamefont {Frank}}, \bibinfo {author} {\bibfnamefont {Tsung-Han}\
  \bibnamefont {Lee}}, \bibinfo {author} {\bibfnamefont {Gargee}\ \bibnamefont
  {Bhattacharyya}}, \bibinfo {author} {\bibfnamefont {Pak Ki~Henry}\
  \bibnamefont {Tsang}}, \bibinfo {author} {\bibfnamefont {Victor~L.}\
  \bibnamefont {Quito}}, \bibinfo {author} {\bibfnamefont {Vladimir}\
  \bibnamefont {Dobrosavljevi\ifmmode~\acute{c}\else \'{c}\fi{}}}, \bibinfo
  {author} {\bibfnamefont {Ove}\ \bibnamefont {Christiansen}}, \ and\ \bibinfo
  {author} {\bibfnamefont {Nicola}\ \bibnamefont {Lanat\`a}},\ }\bibfield
  {title} {\enquote {\bibinfo {title} {Quantum embedding description of the
  anderson lattice model with the ghost gutzwiller approximation},}\ }\href
  {\doibase 10.1103/PhysRevB.104.L081103} {\bibfield  {journal} {\bibinfo
  {journal} {Phys. Rev. B}\ }\textbf {\bibinfo {volume} {104}},\ \bibinfo
  {pages} {L081103} (\bibinfo {year} {2021})}\BibitemShut {NoStop}%
\bibitem [{\citenamefont {Guerci}\ \emph {et~al.}(2019)\citenamefont {Guerci},
  \citenamefont {Capone},\ and\ \citenamefont {Fabrizio}}]{guerci_2019}%
  \BibitemOpen
  \bibfield  {author} {\bibinfo {author} {\bibfnamefont {Daniele}\ \bibnamefont
  {Guerci}}, \bibinfo {author} {\bibfnamefont {Massimo}\ \bibnamefont
  {Capone}}, \ and\ \bibinfo {author} {\bibfnamefont {Michele}\ \bibnamefont
  {Fabrizio}},\ }\bibfield  {title} {\enquote {\bibinfo {title} {Exciton mott
  transition revisited},}\ }\href {\doibase 10.1103/PhysRevMaterials.3.054605}
  {\bibfield  {journal} {\bibinfo  {journal} {Phys. Rev. Materials}\ }\textbf
  {\bibinfo {volume} {3}},\ \bibinfo {pages} {054605} (\bibinfo {year}
  {2019})}\BibitemShut {NoStop}%
\bibitem [{\citenamefont {Guerci}(2019)}]{guerci_phd_2019}%
  \BibitemOpen
  \bibfield  {author} {\bibinfo {author} {\bibfnamefont {Daniele}\ \bibnamefont
  {Guerci}},\ }\emph {\bibinfo {title} {Beyond simple variational approaches to
  strongly correlated electron systems}},\ \href
  {https://iris.sissa.it/handle/20.500.11767/103994#.YZBFIC9w2Wg} {Ph.D.
  thesis},\ \bibinfo  {school} {International School for Advanced Studies
  (SISSA)} (\bibinfo {year} {2019})\BibitemShut {NoStop}%
\bibitem [{\citenamefont {Zhang}\ and\ \citenamefont
  {Sachdev}(2020)}]{ancilla_zhang_2020}%
  \BibitemOpen
  \bibfield  {author} {\bibinfo {author} {\bibfnamefont {Ya-Hui}\ \bibnamefont
  {Zhang}}\ and\ \bibinfo {author} {\bibfnamefont {Subir}\ \bibnamefont
  {Sachdev}},\ }\bibfield  {title} {\enquote {\bibinfo {title} {From the
  pseudogap metal to the fermi liquid using ancilla qubits},}\ }\href {\doibase
  10.1103/PhysRevResearch.2.023172} {\bibfield  {journal} {\bibinfo  {journal}
  {Phys. Rev. Research}\ }\textbf {\bibinfo {volume} {2}},\ \bibinfo {pages}
  {023172} (\bibinfo {year} {2020})}\BibitemShut {NoStop}%
\bibitem [{\citenamefont {Nikolaenko}\ \emph {et~al.}(2021)\citenamefont
  {Nikolaenko}, \citenamefont {Tikhanovskaya}, \citenamefont {Sachdev},\ and\
  \citenamefont {Zhang}}]{ancilla_nikolaenko_2021}%
  \BibitemOpen
  \bibfield  {author} {\bibinfo {author} {\bibfnamefont {Alexander}\
  \bibnamefont {Nikolaenko}}, \bibinfo {author} {\bibfnamefont {Maria}\
  \bibnamefont {Tikhanovskaya}}, \bibinfo {author} {\bibfnamefont {Subir}\
  \bibnamefont {Sachdev}}, \ and\ \bibinfo {author} {\bibfnamefont {Ya-Hui}\
  \bibnamefont {Zhang}},\ }\bibfield  {title} {\enquote {\bibinfo {title}
  {Small to large fermi surface transition in a single-band model using
  randomly coupled ancillas},}\ }\href {\doibase 10.1103/PhysRevB.103.235138}
  {\bibfield  {journal} {\bibinfo  {journal} {Phys. Rev. B}\ }\textbf {\bibinfo
  {volume} {103}},\ \bibinfo {pages} {235138} (\bibinfo {year}
  {2021})}\BibitemShut {NoStop}%
\bibitem [{\citenamefont {Ng}\ and\ \citenamefont
  {Cheng}(1999)}]{slavefermion3}%
  \BibitemOpen
  \bibfield  {author} {\bibinfo {author} {\bibfnamefont {T.~K.}\ \bibnamefont
  {Ng}}\ and\ \bibinfo {author} {\bibfnamefont {C.~H.}\ \bibnamefont {Cheng}},\
  }\bibfield  {title} {\enquote {\bibinfo {title} {Supersymmetry in models with
  strong on-site coulomb repulsion: Application to the heisenberg model},}\
  }\href {\doibase 10.1103/PhysRevB.59.R6616} {\bibfield  {journal} {\bibinfo
  {journal} {Phys. Rev. B}\ }\textbf {\bibinfo {volume} {59}},\ \bibinfo
  {pages} {R6616--R6619} (\bibinfo {year} {1999})}\BibitemShut {NoStop}%
\bibitem [{SI()}]{SI}%
  \BibitemOpen
  \href@noop {} {}\bibinfo {note} {See supplementary information below for
  further details.}\BibitemShut {Stop}%
\bibitem [{\citenamefont {Kwon}\ \emph {et~al.}(1993)\citenamefont {Kwon},
  \citenamefont {Ceperley},\ and\ \citenamefont {Martin}}]{Kwon1993Backflow}%
  \BibitemOpen
  \bibfield  {author} {\bibinfo {author} {\bibfnamefont {Yongkyung}\
  \bibnamefont {Kwon}}, \bibinfo {author} {\bibfnamefont {D.~M.}\ \bibnamefont
  {Ceperley}}, \ and\ \bibinfo {author} {\bibfnamefont {Richard~M.}\
  \bibnamefont {Martin}},\ }\bibfield  {title} {\enquote {\bibinfo {title}
  {Effects of three-body and backflow correlations in the two-dimensional
  electron gas},}\ }\href {\doibase 10.1103/PhysRevB.48.12037} {\bibfield
  {journal} {\bibinfo  {journal} {Phys. Rev. B}\ }\textbf {\bibinfo {volume}
  {48}},\ \bibinfo {pages} {12037--12046} (\bibinfo {year} {1993})}\BibitemShut
  {NoStop}%
\bibitem [{\citenamefont {Kwon}\ \emph {et~al.}(1998)\citenamefont {Kwon},
  \citenamefont {Ceperley},\ and\ \citenamefont {Martin}}]{Kwon1998Backflow}%
  \BibitemOpen
  \bibfield  {author} {\bibinfo {author} {\bibfnamefont {Yongkyung}\
  \bibnamefont {Kwon}}, \bibinfo {author} {\bibfnamefont {D.~M.}\ \bibnamefont
  {Ceperley}}, \ and\ \bibinfo {author} {\bibfnamefont {Richard~M.}\
  \bibnamefont {Martin}},\ }\bibfield  {title} {\enquote {\bibinfo {title}
  {Effects of backflow correlation in the three-dimensional electron gas:
  Quantum monte carlo study},}\ }\href {\doibase 10.1103/PhysRevB.58.6800}
  {\bibfield  {journal} {\bibinfo  {journal} {Phys. Rev. B}\ }\textbf {\bibinfo
  {volume} {58}},\ \bibinfo {pages} {6800--6806} (\bibinfo {year}
  {1998})}\BibitemShut {NoStop}%
\bibitem [{\citenamefont {Tocchio}\ \emph {et~al.}(2008)\citenamefont
  {Tocchio}, \citenamefont {Becca}, \citenamefont {Parola},\ and\ \citenamefont
  {Sorella}}]{Tocchio2008Backflow}%
  \BibitemOpen
  \bibfield  {author} {\bibinfo {author} {\bibfnamefont {Luca~F.}\ \bibnamefont
  {Tocchio}}, \bibinfo {author} {\bibfnamefont {Federico}\ \bibnamefont
  {Becca}}, \bibinfo {author} {\bibfnamefont {Alberto}\ \bibnamefont {Parola}},
  \ and\ \bibinfo {author} {\bibfnamefont {Sandro}\ \bibnamefont {Sorella}},\
  }\bibfield  {title} {\enquote {\bibinfo {title} {Role of backflow
  correlations for the nonmagnetic phase of the
  $t\text{--}{t}^{\ensuremath{'}}$ hubbard model},}\ }\href {\doibase
  10.1103/PhysRevB.78.041101} {\bibfield  {journal} {\bibinfo  {journal} {Phys.
  Rev. B}\ }\textbf {\bibinfo {volume} {78}},\ \bibinfo {pages} {041101}
  (\bibinfo {year} {2008})}\BibitemShut {NoStop}%
\bibitem [{\citenamefont {Tocchio}\ \emph {et~al.}(2011)\citenamefont
  {Tocchio}, \citenamefont {Becca},\ and\ \citenamefont
  {Gros}}]{Tocchio2011Backflow}%
  \BibitemOpen
  \bibfield  {author} {\bibinfo {author} {\bibfnamefont {Luca~F.}\ \bibnamefont
  {Tocchio}}, \bibinfo {author} {\bibfnamefont {Federico}\ \bibnamefont
  {Becca}}, \ and\ \bibinfo {author} {\bibfnamefont {Claudius}\ \bibnamefont
  {Gros}},\ }\bibfield  {title} {\enquote {\bibinfo {title} {Backflow
  correlations in the hubbard model: An efficient tool for the study of the
  metal-insulator transition and the large-$u$ limit},}\ }\href {\doibase
  10.1103/PhysRevB.83.195138} {\bibfield  {journal} {\bibinfo  {journal} {Phys.
  Rev. B}\ }\textbf {\bibinfo {volume} {83}},\ \bibinfo {pages} {195138}
  (\bibinfo {year} {2011})}\BibitemShut {NoStop}%
\bibitem [{\citenamefont {Cybenko}(1989)}]{Cybenko1989universal}%
  \BibitemOpen
  \bibfield  {author} {\bibinfo {author} {\bibfnamefont {G.}~\bibnamefont
  {Cybenko}},\ }\bibfield  {title} {\enquote {\bibinfo {title} {Approximation
  by superpositions of a sigmoidal function},}\ }\href {\doibase
  10.1007/BF02551274} {\bibfield  {journal} {\bibinfo  {journal} {Mathematics
  of Control, Signals and Systems}\ }\textbf {\bibinfo {volume} {2}},\ \bibinfo
  {pages} {303--314} (\bibinfo {year} {1989})}\BibitemShut {NoStop}%
\bibitem [{\citenamefont {Zhao}\ \emph {et~al.}(2017)\citenamefont {Zhao},
  \citenamefont {Ido}, \citenamefont {Morita},\ and\ \citenamefont
  {Imada}}]{Hui-Hai2017treeTN}%
  \BibitemOpen
  \bibfield  {author} {\bibinfo {author} {\bibfnamefont {Hui-Hai}\ \bibnamefont
  {Zhao}}, \bibinfo {author} {\bibfnamefont {Kota}\ \bibnamefont {Ido}},
  \bibinfo {author} {\bibfnamefont {Satoshi}\ \bibnamefont {Morita}}, \ and\
  \bibinfo {author} {\bibfnamefont {Masatoshi}\ \bibnamefont {Imada}},\
  }\bibfield  {title} {\enquote {\bibinfo {title} {Variational monte carlo
  method for fermionic models combined with tensor networks and applications to
  the hole-doped two-dimensional hubbard model},}\ }\href {\doibase
  10.1103/PhysRevB.96.085103} {\bibfield  {journal} {\bibinfo  {journal} {Phys.
  Rev. B}\ }\textbf {\bibinfo {volume} {96}},\ \bibinfo {pages} {085103}
  (\bibinfo {year} {2017})}\BibitemShut {NoStop}%
\bibitem [{\citenamefont {Sorella}\ \emph {et~al.}(2007)\citenamefont
  {Sorella}, \citenamefont {Casula},\ and\ \citenamefont
  {Rocca}}]{sorella2007StochasticReconfiguration}%
  \BibitemOpen
  \bibfield  {author} {\bibinfo {author} {\bibfnamefont {Sandro}\ \bibnamefont
  {Sorella}}, \bibinfo {author} {\bibfnamefont {Michele}\ \bibnamefont
  {Casula}}, \ and\ \bibinfo {author} {\bibfnamefont {Dario}\ \bibnamefont
  {Rocca}},\ }\bibfield  {title} {\enquote {\bibinfo {title} {Weak binding
  between two aromatic rings: Feeling the van der waals attraction by quantum
  monte carlo methods},}\ }\href {\doibase 10.1063/1.2746035} {\bibfield
  {journal} {\bibinfo  {journal} {The Journal of Chemical Physics}\ }\textbf
  {\bibinfo {volume} {127}},\ \bibinfo {pages} {014105} (\bibinfo {year}
  {2007})},\ \Eprint {http://arxiv.org/abs/https://doi.org/10.1063/1.2746035}
  {https://doi.org/10.1063/1.2746035} \BibitemShut {NoStop}%
\bibitem [{\citenamefont {Amari}(1998)}]{amari1998natural}%
  \BibitemOpen
  \bibfield  {author} {\bibinfo {author} {\bibfnamefont {Shun-Ichi}\
  \bibnamefont {Amari}},\ }\bibfield  {title} {\enquote {\bibinfo {title}
  {Natural gradient works efficiently in learning},}\ }\href@noop {} {\bibfield
   {journal} {\bibinfo  {journal} {Neural computation}\ }\textbf {\bibinfo
  {volume} {10}},\ \bibinfo {pages} {251--276} (\bibinfo {year}
  {1998})}\BibitemShut {NoStop}%
\bibitem [{\citenamefont {Carleo}\ \emph {et~al.}(2019)\citenamefont {Carleo},
  \citenamefont {Choo}, \citenamefont {Hofmann}, \citenamefont {Smith},
  \citenamefont {Westerhout}, \citenamefont {Alet}, \citenamefont {Davis},
  \citenamefont {Efthymiou}, \citenamefont {Glasser}, \citenamefont {Lin},
  \citenamefont {Mauri}, \citenamefont {Mazzola}, \citenamefont {Mendl},
  \citenamefont {van Nieuwenburg}, \citenamefont {O'Reilly}, \citenamefont
  {Th{\'e}veniaut}, \citenamefont {Torlai}, \citenamefont {Vicentini},\ and\
  \citenamefont {Wietek}}]{netket:2019}%
  \BibitemOpen
  \bibfield  {author} {\bibinfo {author} {\bibfnamefont {Giuseppe}\
  \bibnamefont {Carleo}}, \bibinfo {author} {\bibfnamefont {Kenny}\
  \bibnamefont {Choo}}, \bibinfo {author} {\bibfnamefont {Damian}\ \bibnamefont
  {Hofmann}}, \bibinfo {author} {\bibfnamefont {James E.~T.}\ \bibnamefont
  {Smith}}, \bibinfo {author} {\bibfnamefont {Tom}\ \bibnamefont {Westerhout}},
  \bibinfo {author} {\bibfnamefont {Fabien}\ \bibnamefont {Alet}}, \bibinfo
  {author} {\bibfnamefont {Emily~J.}\ \bibnamefont {Davis}}, \bibinfo {author}
  {\bibfnamefont {Stavros}\ \bibnamefont {Efthymiou}}, \bibinfo {author}
  {\bibfnamefont {Ivan}\ \bibnamefont {Glasser}}, \bibinfo {author}
  {\bibfnamefont {Sheng-Hsuan}\ \bibnamefont {Lin}}, \bibinfo {author}
  {\bibfnamefont {Marta}\ \bibnamefont {Mauri}}, \bibinfo {author}
  {\bibfnamefont {Guglielmo}\ \bibnamefont {Mazzola}}, \bibinfo {author}
  {\bibfnamefont {Christian~B.}\ \bibnamefont {Mendl}}, \bibinfo {author}
  {\bibfnamefont {Evert}\ \bibnamefont {van Nieuwenburg}}, \bibinfo {author}
  {\bibfnamefont {Ossian}\ \bibnamefont {O'Reilly}}, \bibinfo {author}
  {\bibfnamefont {Hugo}\ \bibnamefont {Th{\'e}veniaut}}, \bibinfo {author}
  {\bibfnamefont {Giacomo}\ \bibnamefont {Torlai}}, \bibinfo {author}
  {\bibfnamefont {Filippo}\ \bibnamefont {Vicentini}}, \ and\ \bibinfo {author}
  {\bibfnamefont {Alexander}\ \bibnamefont {Wietek}},\ }\bibfield  {title}
  {\enquote {\bibinfo {title} {Netket: A machine learning toolkit for many-body
  quantum systems},}\ }\href {\doibase 10.1016/j.softx.2019.100311} {\bibfield
  {journal} {\bibinfo  {journal} {SoftwareX}\ ,\ \bibinfo {pages} {100311}}
  (\bibinfo {year} {2019})}\BibitemShut {NoStop}%
\bibitem [{\citenamefont {Bradbury}\ \emph {et~al.}(2018)\citenamefont
  {Bradbury}, \citenamefont {Frostig}, \citenamefont {Hawkins}, \citenamefont
  {Johnson}, \citenamefont {Leary}, \citenamefont {Maclaurin}, \citenamefont
  {Necula}, \citenamefont {Paszke}, \citenamefont {Vander{P}las}, \citenamefont
  {Wanderman-{M}ilne},\ and\ \citenamefont {Zhang}}]{jax2018github}%
  \BibitemOpen
  \bibfield  {author} {\bibinfo {author} {\bibfnamefont {James}\ \bibnamefont
  {Bradbury}}, \bibinfo {author} {\bibfnamefont {Roy}\ \bibnamefont {Frostig}},
  \bibinfo {author} {\bibfnamefont {Peter}\ \bibnamefont {Hawkins}}, \bibinfo
  {author} {\bibfnamefont {Matthew~James}\ \bibnamefont {Johnson}}, \bibinfo
  {author} {\bibfnamefont {Chris}\ \bibnamefont {Leary}}, \bibinfo {author}
  {\bibfnamefont {Dougal}\ \bibnamefont {Maclaurin}}, \bibinfo {author}
  {\bibfnamefont {George}\ \bibnamefont {Necula}}, \bibinfo {author}
  {\bibfnamefont {Adam}\ \bibnamefont {Paszke}}, \bibinfo {author}
  {\bibfnamefont {Jake}\ \bibnamefont {Vander{P}las}}, \bibinfo {author}
  {\bibfnamefont {Skye}\ \bibnamefont {Wanderman-{M}ilne}}, \ and\ \bibinfo
  {author} {\bibfnamefont {Qiao}\ \bibnamefont {Zhang}},\ }\href
  {http://github.com/google/jax} {\enquote {\bibinfo {title} {{JAX}: composable
  transformations of {P}ython+{N}um{P}y programs},}\ } (\bibinfo {year}
  {2018})\BibitemShut {NoStop}%
\bibitem [{\citenamefont {Kashima}\ and\ \citenamefont
  {Imada}(2001)}]{Kashima2001varianceExtrapolation}%
  \BibitemOpen
  \bibfield  {author} {\bibinfo {author} {\bibfnamefont {Tsuyoshi}\
  \bibnamefont {Kashima}}\ and\ \bibinfo {author} {\bibfnamefont {Masatoshi}\
  \bibnamefont {Imada}},\ }\bibfield  {title} {\enquote {\bibinfo {title}
  {Path-integral renormalization group method for numerical study on ground
  states of strongly correlated electronic systems},}\ }\href {\doibase
  10.1143/JPSJ.70.2287} {\bibfield  {journal} {\bibinfo  {journal} {Journal of
  the Physical Society of Japan}\ }\textbf {\bibinfo {volume} {70}},\ \bibinfo
  {pages} {2287--2299} (\bibinfo {year} {2001})},\ \Eprint
  {http://arxiv.org/abs/https://doi.org/10.1143/JPSJ.70.2287}
  {https://doi.org/10.1143/JPSJ.70.2287} \BibitemShut {NoStop}%
\bibitem [{\citenamefont {Montúfar}\ \emph {et~al.}(2014)\citenamefont
  {Montúfar}, \citenamefont {Pascanu}, \citenamefont {Cho},\ and\
  \citenamefont {Bengio}}]{montufar2014deepernets}%
  \BibitemOpen
  \bibfield  {author} {\bibinfo {author} {\bibfnamefont {Guido}\ \bibnamefont
  {Montúfar}}, \bibinfo {author} {\bibfnamefont {Razvan}\ \bibnamefont
  {Pascanu}}, \bibinfo {author} {\bibfnamefont {Kyunghyun}\ \bibnamefont
  {Cho}}, \ and\ \bibinfo {author} {\bibfnamefont {Yoshua}\ \bibnamefont
  {Bengio}},\ }\href@noop {} {\enquote {\bibinfo {title} {On the number of
  linear regions of deep neural networks},}\ } (\bibinfo {year} {2014}),\
  \Eprint {http://arxiv.org/abs/1402.1869} {arXiv:1402.1869 [stat.ML]}
  \BibitemShut {NoStop}%
\bibitem [{\citenamefont {Zheng}\ \emph {et~al.}(2017)\citenamefont {Zheng},
  \citenamefont {Chung}, \citenamefont {Corboz}, \citenamefont {Ehlers},
  \citenamefont {Qin}, \citenamefont {Noack}, \citenamefont {Shi},
  \citenamefont {White}, \citenamefont {Zhang},\ and\ \citenamefont
  {Chan}}]{Zheng2017stripes}%
  \BibitemOpen
  \bibfield  {author} {\bibinfo {author} {\bibfnamefont {Bo-Xiao}\ \bibnamefont
  {Zheng}}, \bibinfo {author} {\bibfnamefont {Chia-Min}\ \bibnamefont {Chung}},
  \bibinfo {author} {\bibfnamefont {Philippe}\ \bibnamefont {Corboz}}, \bibinfo
  {author} {\bibfnamefont {Georg}\ \bibnamefont {Ehlers}}, \bibinfo {author}
  {\bibfnamefont {Ming-Pu}\ \bibnamefont {Qin}}, \bibinfo {author}
  {\bibfnamefont {Reinhard~M.}\ \bibnamefont {Noack}}, \bibinfo {author}
  {\bibfnamefont {Hao}\ \bibnamefont {Shi}}, \bibinfo {author} {\bibfnamefont
  {Steven~R.}\ \bibnamefont {White}}, \bibinfo {author} {\bibfnamefont
  {Shiwei}\ \bibnamefont {Zhang}}, \ and\ \bibinfo {author} {\bibfnamefont
  {Garnet Kin-Lic}\ \bibnamefont {Chan}},\ }\bibfield  {title} {\enquote
  {\bibinfo {title} {Stripe order in the underdoped region of the
  two-dimensional hubbard model},}\ }\href {\doibase 10.1126/science.aam7127}
  {\bibfield  {journal} {\bibinfo  {journal} {Science}\ }\textbf {\bibinfo
  {volume} {358}},\ \bibinfo {pages} {1155--1160} (\bibinfo {year} {2017})},\
  \Eprint
  {http://arxiv.org/abs/https://science.sciencemag.org/content/358/6367/1155.full.pdf}
  {https://science.sciencemag.org/content/358/6367/1155.full.pdf} \BibitemShut
  {NoStop}%
\bibitem [{\citenamefont {Dagotto}\ \emph {et~al.}(1992)\citenamefont
  {Dagotto}, \citenamefont {Moreo}, \citenamefont {Ortolani}, \citenamefont
  {Poilblanc},\ and\ \citenamefont {Riera}}]{Dagotto1992ED}%
  \BibitemOpen
  \bibfield  {author} {\bibinfo {author} {\bibfnamefont {E.}~\bibnamefont
  {Dagotto}}, \bibinfo {author} {\bibfnamefont {A.}~\bibnamefont {Moreo}},
  \bibinfo {author} {\bibfnamefont {F.}~\bibnamefont {Ortolani}}, \bibinfo
  {author} {\bibfnamefont {D.}~\bibnamefont {Poilblanc}}, \ and\ \bibinfo
  {author} {\bibfnamefont {J.}~\bibnamefont {Riera}},\ }\bibfield  {title}
  {\enquote {\bibinfo {title} {Static and dynamical properties of doped hubbard
  clusters},}\ }\href {\doibase 10.1103/PhysRevB.45.10741} {\bibfield
  {journal} {\bibinfo  {journal} {Phys. Rev. B}\ }\textbf {\bibinfo {volume}
  {45}},\ \bibinfo {pages} {10741--10760} (\bibinfo {year} {1992})}\BibitemShut
  {NoStop}%
\bibitem [{\citenamefont {Qin}\ \emph {et~al.}(2016)\citenamefont {Qin},
  \citenamefont {Shi},\ and\ \citenamefont {Zhang}}]{Qin2026AFQMCBenchmark}%
  \BibitemOpen
  \bibfield  {author} {\bibinfo {author} {\bibfnamefont {Mingpu}\ \bibnamefont
  {Qin}}, \bibinfo {author} {\bibfnamefont {Hao}\ \bibnamefont {Shi}}, \ and\
  \bibinfo {author} {\bibfnamefont {Shiwei}\ \bibnamefont {Zhang}},\ }\bibfield
   {title} {\enquote {\bibinfo {title} {Benchmark study of the two-dimensional
  hubbard model with auxiliary-field quantum monte carlo method},}\ }\href
  {\doibase 10.1103/PhysRevB.94.085103} {\bibfield  {journal} {\bibinfo
  {journal} {Phys. Rev. B}\ }\textbf {\bibinfo {volume} {94}},\ \bibinfo
  {pages} {085103} (\bibinfo {year} {2016})}\BibitemShut {NoStop}%
\bibitem [{\citenamefont {Wietek}\ \emph {et~al.}(2021)\citenamefont {Wietek},
  \citenamefont {He}, \citenamefont {White}, \citenamefont {Georges},\ and\
  \citenamefont {Stoudenmire}}]{Wietek2021stripes}%
  \BibitemOpen
  \bibfield  {author} {\bibinfo {author} {\bibfnamefont {Alexander}\
  \bibnamefont {Wietek}}, \bibinfo {author} {\bibfnamefont {Yuan-Yao}\
  \bibnamefont {He}}, \bibinfo {author} {\bibfnamefont {Steven~R.}\
  \bibnamefont {White}}, \bibinfo {author} {\bibfnamefont {Antoine}\
  \bibnamefont {Georges}}, \ and\ \bibinfo {author} {\bibfnamefont {E.~Miles}\
  \bibnamefont {Stoudenmire}},\ }\bibfield  {title} {\enquote {\bibinfo {title}
  {Stripes, antiferromagnetism, and the pseudogap in the doped hubbard model at
  finite temperature},}\ }\href {\doibase 10.1103/PhysRevX.11.031007}
  {\bibfield  {journal} {\bibinfo  {journal} {Phys. Rev. X}\ }\textbf {\bibinfo
  {volume} {11}},\ \bibinfo {pages} {031007} (\bibinfo {year}
  {2021})}\BibitemShut {NoStop}%
\bibitem [{\citenamefont {Chowdhury}\ \emph {et~al.}(2021)\citenamefont
  {Chowdhury}, \citenamefont {Georges}, \citenamefont {Parcollet},\ and\
  \citenamefont {Sachdev}}]{SYK_review}%
  \BibitemOpen
  \bibfield  {author} {\bibinfo {author} {\bibfnamefont {Debanjan}\
  \bibnamefont {Chowdhury}}, \bibinfo {author} {\bibfnamefont {Antoine}\
  \bibnamefont {Georges}}, \bibinfo {author} {\bibfnamefont {Olivier}\
  \bibnamefont {Parcollet}}, \ and\ \bibinfo {author} {\bibfnamefont {Subir}\
  \bibnamefont {Sachdev}},\ }\href {\doibase 10.48550/ARXIV.2109.05037}
  {\enquote {\bibinfo {title} {Sachdev-ye-kitaev models and beyond: A window
  into non-fermi liquids},}\ } (\bibinfo {year} {2021})\BibitemShut {NoStop}%
\end{thebibliography}%

\onecolumngrid
\vspace{10pt}
\appendix
\section{Universality of the hidden fermion determinant state in the lattice}\label{Apndx A: universality}
The aim of this appendix is to prove that the hidden fermion determinant state of amplitudes $\psi (x)$, can exactly match the combinatorially-many amplitudes of an arbitrary target state $\psi_{\rm tar}(x)$, thus proving $\psi (x)$ to be a universal wave function {\it ansatz} in the fixed particle subspace of Fock space.

Take $N$ and $\tilde{N}$ visible and hidden fermions occupying  $M$ and $\tilde{M}$ visible and hidden modes respectively. For simplicity, and without loss of generality, we consider the particular case where the hidden orbitals have zero amplitude in the visible lattice and the visible orbitals have zero amplitude in the positions of the hidden modes, i.e. $\chi_{\rm v} = \phi_{\rm h} = 0$, thus simplifying the evaluation of the amplitudes of the hidden fermion determinant state to:
\begin{equation}
    \psi(x) = \det
    \begin{bmatrix}
    \phi_{\rm v}(x) & 0 \\
    0 & \chi_{\rm h}(f(x))
    \end{bmatrix}
     = \det
     \begin{bmatrix}
     \phi_{\rm v}(x)
     \end{bmatrix}
     \cdot \det
     \begin{bmatrix}
     \chi_{\rm h}(f(x))
     \end{bmatrix}.
\end{equation}
With this particular wave function structure, we already know that $\det \big[\phi_{\rm v}(x) \big]$ is not capable of matching the amplitudes of an arbitrary state in Fock space. This is because it is a $N\times N$ determinant of single-particle orbitals, thus incapable of capturing correlated states. This implies that the majority of the expressive power of the {\it ansatz} must come from the $\tilde{N}\times \tilde{N}$ determinant $\det \big[ \chi_{\rm h}(f(x))\big]$. Note that, while the $\chi_{\rm h}$ are single particle orbitals of the hidden particle positions, the hidden particle configuration depends on the configuration of all of the visible particles through the constraint function, and consequently, $\det \big[ \chi_{\rm h}(f(x))\big]$ can be interpreted as the determinant of multiple-visible-particle orbitals. Therefore, in order to have $\psi(x) = \psi_{\rm tar}(x)$, the determinant $\det \big[ \chi_{\rm h}(f(x))\big]$ must satisfy:
\begin{equation}\label{eq: universal req}
    \det \big[ \chi_{\rm h}(f(x))\big] = \frac{\psi_{\rm tar}(x)}{\det\big[\phi_{\rm v}(x)\big]},
\end{equation}
 which can only be accomplished iff $\det[\phi_{\rm v}(x)]\neq 0$ for all $ x $. This requirement is satisfied by choosing the rows of the $(M \times N)$ $\phi_{\rm v}$ matrix to be non-colinear $N$-dimenstional vectors. Note that both $\psi_{\rm tar}(x)$ and $\det\big[\phi_{\rm v}(x)\big]$ are anti-symmetric functions, and in consequence, $\det \big[ \chi_{\rm h}(f(x))\big]$ must be a symmetric function of $x$, which is achieved by taking $f(x)$ to be symmetric.
 
 Now we provide an explicit construction of $f(x)$ and $\chi_{\rm h}$ such that requirement~\ref{eq: universal req} is satisfied for an arbitrary state $\psi_{\rm tar}(x)$. This construction is based on identifying $\det \big[ \chi_{\rm h}(f(x))\big]$ with a lookup table with $\alpha_{\rm max} = \binom{M}{N}$ entries, that provides the corresponding amplitudes for the different $\alpha_{\rm max}$ occupation configurations $n$ of the visible particles on the visible modes. Therefore, in the worst case scenario we require $\tilde{M} = \alpha_{\rm max} + (\tilde{N}-1)$ hidden modes. A possible choice for the $\left( \tilde{M} \times \tilde{N}\right)$ matrix of amplitudes $\chi_{\rm h}$ that achieves the desired lookup table is given by
 \begin{equation}
     \chi_{\rm h} = \begin{bmatrix}
     \chi_1 & 0 & \hdots      &0 \\
     \vdots  & \vdots  &\;      &\vdots \\
     \chi_{\alpha_{\rm max}} & 0 &\hdots &0 \\
     0        & \; &\;       &\;  \\   
     \vdots         & \; & \mathbb{I}_{(\tilde{N}-1)}      &\;  \\  
     0         & \; &\;      &  \\  
     \end{bmatrix}
 \end{equation}
 with $\mathbb{I}_{(\tilde{N}-1)}$ the $(\tilde{N}-1) \times (\tilde{N}-1)$ identity matrix. We identify $\chi_1 , ..., \chi_{\alpha_{\rm max}}$ with the $\alpha_{\rm max}$ different amplitudes of the ratio $\psi_{\rm tar}(x)/\det\big[\phi_{\rm v}(x)\big]$. The position of hidden fermion $i$ is determined by $i^{\rm th}$ component of the constraint function, i.e. $f_i(x)$, which must satisfy $0< f_i(x) \leq \tilde{M}$ and $f_i(x) \in \mathbb{Z}$. We choose
 \begin{equation}
     f_i(x) = 
     \begin{cases}
      \alpha(x) & \textrm{if } \; i = 1 \\
      \alpha_{\rm max} + (i-1)  & \textrm{if } \; 2 \leq i\leq \tilde{N}
     \end{cases}
 \end{equation}
 where $0<\alpha (x) \leq\alpha_{\rm max}$ and $\alpha \in \mathbb{Z}$ provides a distinct label to each of the $\binom{M}{N}$ different occupation configurations of the visible particles on the visible modes.
 
 With the construction we obtain $\det \big[ \chi_{\rm h}(f(x))\big] = \chi_m = \psi_{\rm tar}(x)/\det\big[\phi_{\rm v}(x)\big]$ with $1\leq m \leq \alpha_{\rm max}$, thus concluding the proof that the hidden fermion determinant state can match the amplitudes of an arbitrary state in the fixed particle subspace of Fock space.

It is noteworthy that the explicit construction of a combinatorially-large number of hidden modes can be circumvented by directly parametrizing the function composition $\big[\chi_{\rm h}(f(x))\big]$ by a universal function approximator.

\section{Configuration interaction wavefunctions from the hidden fermion determinant state}\label{Apndx A1: CI wavefunction}
The aim of this appendix is to provide an explicit construction of a multi-determinant wave function ansatz from the hidden fermion determinant state. In particular, we show that linear combinations of the Hartree-Fock (HF) ground state with all possible single, double, triple,.., $\tilde{N}$-tuple excitations to the $\tilde{N}$ lower virtual orbitals can be constructed from the hidden fermion determinant state with $\tilde{N}$ hidden fermions. First we introduce the notation that will be used in the remaining of the section. Then we show the $\tilde{N} = 1$ and $\tilde{N} = 2$ cases to demonstrate that they lead to a linear combination of $N$-particle determinants containing single and double excitations to the first and second virtual orbitals respectively. We conclude by generalizing the result to an arbitrary number of hidden fermions.

 $\{\phi_\alpha^{\rm HF}\}$ is the set of single-particle the HF orbitals. The subscript $\alpha = 1, 2, ..., N_{\rm tot}$ labels the orbitals in ascending order in energy. We refer to the $N$ lowest energy orbitals as the occupied orbitals. The remaining orbitals are called virtual orbitals, following the quantum chemistry nomenclature. The Slater determinant constructed with  $\phi_1^{\rm HF}, ..., \phi_N^{\rm HF}$ corresponds to the usual HF ground state $|\Phi\rangle = \hat{\varphi}^\dag_1\hdots \hat{\varphi}^\dag_N |0\rangle$, whose amplitudes are given by the determinant:
\begin{equation}
    \psi^{\rm HF}(x) = 
    \det
    \begin{bmatrix}
    \phi_{1}^{\rm HF}(x_1)& \cdots   & \phi_{N}^{\rm HF}(x_{1}) \\
    \vdots&    & \vdots   \\
    \phi_{1}^{\rm HF}(x_{N})& \cdots   & \phi_{N}^{\rm HF}(x_{N}))\\
    \end{bmatrix}.    
\end{equation}
Following the Quantum Chemistry notation $|\Phi_\alpha^\mu \rangle$ with $\alpha\leq N$ and $\mu>N$, is the excitation of the fermion in occupied orbital $\alpha$ to the virtual orbital $\mu$:
\begin{equation}
    |\Phi_\alpha^\mu \rangle = \hat{\varphi}^\dag_1 \hdots \hat{\varphi}^\dag_{\alpha-1}\hat{\varphi}^\dag_{\alpha+1}\hdots\hat{\varphi}^\dag_{N} \cdot \hat{\varphi}^\dag_{\mu} |0\rangle.
\end{equation}
Analogously $|\Phi_{\alpha_1\hdots\alpha_n}^{\mu_1\hdots\mu_n} \rangle$ labels the promotion of the $n$ electrons in occupied orbitals $\alpha_1$ through $\alpha_n$ to virtual orbitals $\mu_1$ through $\mu_n$.

We choose to identify the visible orbitals $\phi_{\rm v}$ with the $N$ occupied HF orbitals $\phi_\alpha^{\rm HF}$ ($1\leq\alpha\leq N$) and the $\tilde{N}$ hidden orbitals $\chi_{\rm h}$ with the $\tilde{N}$ first virtual orbitals$\phi_\mu^{\rm HF}$ ($N<\mu\leq N+\tilde{N}$).

\subsection{$\tilde{N} = 1$ and single excitations}
$\;$

According to the above identification of orbitals, the wave function amplitudes become:
\begin{equation}
    \psi(x) = 
    \det
    \begin{bmatrix}
    \phi_{1}^{\rm HF}(x_1)& \cdots   & \phi_{N}^{\rm HF}(x_{1})& \phi_{N+1}^{\rm HF}(x_{1}) \\
    \vdots&    & \vdots &\vdots  \\
    \phi_{1}^{\rm HF}(x_{N})& \cdots   & \phi_{N}^{\rm HF}(x_{N})& \phi_{N+1}^{\rm HF}(x_{\rm N}))\\
    \;&    &   &  \\
    \hline
    \;&    &   &  \\
    \phi_{1}^{\rm HF}(f_1(x))& \cdots   & \phi_{N}^{\rm HF}(f_1(x))& \phi_{N+1}^{\rm HF}(f_1(x)) \\
    \end{bmatrix}. 
\end{equation}
The cofactor of the above determinant $C_{ij}(x)$ is defined as $C_{ij}(x) = \det [A_{ij}(x)]$, where $A_{ij}(x)$ is the minor of the matrix of orbitals, obtained by eliminating row $i$ and column $j$ from the matrix that enters the original determinant. Using the Laplace expansion of the determinant along its last row:
\begin{equation}
    \psi(x) = \sum_{1\leq \alpha \leq N_{\rm{tot}}} (-1)^{\alpha}\cdot \phi^{\rm HF}_\alpha(f_1(x))\cdot C_{N_{\rm {tot}}, \alpha}(x).
\end{equation}
We realize that $C_{N_{\rm {tot}}, \alpha}(x)$ can be identified with the amplitudes of the $N$-particle Slater determinant obtained from promoting particle in occupied orbital $\alpha$ to the first virtual orbital. In the particular case where $\alpha = N_{\rm tot}$, $C_{N_{\rm {tot}}, N_{\rm {tot}}}(x)$ are the amplitudes of the HF ground state. The prefactors to the cofactors can be set to constants $(-1)^{\alpha}\cdot \phi^{\rm HF}_\alpha(f_1(x)) = c_{\alpha}^{N+1}$. For this choice of constraint function and orbital amplitudes, the projected augmented determinant to the physical space corresponds to the CI wavefunction with single-particle excitations:
\begin{equation}
    |\psi^{\rm proj} \rangle = c_0 |\Phi\rangle + \sum_{1\leq \alpha\leq N}c_{\alpha}^{N+1} |\Phi_{\alpha}^{N+1}\rangle.
\end{equation}

\subsection{$\tilde{N} = 2$ and double excitations}
$\;$

In this case, the amplitudes of the hidden fermion determinant state are given by the determinant:
\begin{equation}\label{eq: 2-excitation determinant}
    \psi(x) = 
    \det
    \begin{bmatrix}
    \phi_{1}^{\rm HF}(x_1)& \cdots   & \phi_{N}^{\rm HF}(x_{1})& \phi_{N+1}^{\rm HF}(x_{1}) & \phi_{N+2}^{\rm HF}(x_{1}) \\
    \vdots&    & \vdots &\vdots  & \vdots\\
    \phi_{1}^{\rm HF}(x_{N})& \cdots   & \phi_{N}^{\rm HF}(x_{N})& \phi_{N+1}^{\rm HF}(x_{\rm N})) & \phi_{N+2}^{\rm HF}(x_{\rm N}))\\
    \;&    &   &  & \\
    \hline
    \;&    &   &  & \\
    \phi_{1}^{\rm HF}(f_1(x))& \cdots   & \phi_{N}^{\rm HF}(f_1(x))& \phi_{N+1}^{\rm HF}(f_1(x)) & \phi_{N+2}^{\rm HF}(f_1(x))\\
    \;&    &   &  & \\
    \phi_{1}^{\rm HF}(f_2(x))& \cdots   & \phi_{N}^{\rm HF}(f_2(x))& \phi_{N+1}^{\rm HF}(f_2(x)) & \phi_{N+2}^{\rm HF}(f_2(x))
    \end{bmatrix}.    
\end{equation}
We now proceed to expand the above determinant on its cofactors by its last row, obtaining the linear combinations of $(N+1)$-particle determinants:
\begin{equation}
    \psi(x) = \sum_{1\leq\alpha\leq N_{\rm tot}} (-1)^\alpha \cdot \phi^{\rm HF}_\alpha(f_2(x))\cdot C_{N_{\rm tot},\alpha}.
\end{equation}
 $C_{i,j}$ is the cofactor of the $(N+2)$-particle determinant $C_{ij} = \det\big[A_{ij} \big]$ where $A_{ij}$ is the minor obtained by removing row $i$ and column $j$ from the matrix entering the determinant in Eq.~\ref{eq: 2-excitation determinant}. Now $C_{N_{\rm tot},\alpha}$ are $(N+1)$-particle determinants. The cofactors $C_{N_{\rm tot},\alpha}$ can also be expanded on their cofactors by their last row, obtaining:
 \begin{equation}
    \psi(x) =  \sum_{\substack{1\leq\alpha\leq N_{\rm tot} \\ \substack{1\leq\beta\leq N_{\rm tot}-1}}} (-1)^{\alpha+\beta+1} \cdot \phi^{\rm HF}_\alpha(f_2(x)) \cdot \phi^{\rm HF}_\beta(f_1(x))\cdot C_{(N_{\rm tot}-1,\beta)(N_{\rm tot},\alpha)}.
\end{equation}
Here $C_{(kl)(ij)}$ is the cofactor $C_{(kl)(ij)} = \det\big[A_{(kl)(ij)} \big]$ where $A_{(kl)(ij)}$ is the minor obtained by removing row $k$ and column $l$ from the previously defined minor $A_{ij}$. Now $ C_{(N_{\rm tot}-1,\beta)(N_{\rm tot},\alpha)}$ are $N$-particle Slater determinants obtained by promoting the particles in the occupied orbitals $\alpha$ and $\beta$ to the two lowest virtual orbitals $\phi^{\rm HF}_{N+1}$ and $\phi^{\rm HF}_{N+2}$. It must be noted that the limiting cases $C_{(N_{\rm tot}-1,N_{\rm tot}-1)(N_{\rm tot},N_{\rm tot})}$, $C_{(N_{\rm tot}-1,N_{\rm tot}-1)(N_{\rm tot},\alpha)}$ and $C_{(N_{\rm tot}-1,\beta)(N_{\rm tot},N_{\rm tot})}$ are the HF ground-state determinant, the excitation of the particle in occupied orbital $\alpha$ to the second virtual orbital and the excitation of the particle in occupied orbital $\beta$ to the first virtual orbital receptively. As before, setting the prefactors to the cofactors to be the constants:
\begin{equation}
    (-1)^{\alpha+\beta+1} \cdot \phi^{\rm HF}_\alpha(f_2(x)) \cdot \phi^{\rm HF}_\alpha(f_2(x))  := 
    \begin{cases}
    c_0 & \textrm{if }  \alpha = N_{\rm tot};\; \beta = N_{\rm tot}-1 \\
    c_\beta^{N+1} & \textrm{if }  \alpha = N_{\rm tot};\; \beta \leq N \\
    c_\alpha^{N+2} & \textrm{if }  \alpha \leq N;\; \beta = N_{\rm tot}-1 \\
    c_{\alpha,\beta}^{N+2, N+1} & \textrm{if }  \alpha \leq N;\; \beta = \leq N
    \end{cases}
\end{equation}
the augmented determinant, projected to the physical space corresponds to the the CI wavefunction with single and double-particle excitations:
\begin{equation}
    |\psi^{\rm proj} \rangle = c_0 |\Psi\rangle + \underbrace{\sum_{ \substack{1\leq \alpha\leq N \\ N < \mu \leq N_{\rm tot}} } c_\alpha^\mu \cdot |\Phi_{\alpha}^\mu \rangle}_{\rm singles} + \underbrace{\sum_{ \substack{ 1\leq \alpha_1 < \alpha_2 \leq N \\ N< \mu_1 < \mu_2\leq N_{\rm tot}} } c_{\alpha_1, \alpha_2}^{\mu_1, \mu_2} \cdot |\Phi_{\alpha_1, \alpha_2}^{\mu_1, \mu_2}\rangle}_{\rm doubles}.
\end{equation}
\subsection{General $\tilde{N}$ and $\tilde{N}$-tuple excitations}
$\;$

The generalization of the two cases above to arbitrary $\tilde{N}$ implies that the hidden fermion determinant state of $\tilde{N}$ hidden fermions can be chosen to be a compact representation of a CI wavefunction with all possible single through $\tilde{N}$-tuple particle excitations to the $\tilde{N}$ lowest virtual orbitals:
\begin{equation}
    |\psi^{\rm proj}\rangle = c_0 \cdot |\Phi \rangle 
    + \underbrace{\sum_{\substack{1\leq \alpha \leq N \\ N< \mu \leq N_{\rm tot}}} c^{\mu}_\alpha|\Phi_\alpha^{\mu}\rangle }_{\rm singles} 
    +\underbrace{\sum_{ \substack{ 1\leq \alpha_1 < \alpha_2 \leq N \\ N< \mu_1 < \mu_2\leq N_{\rm tot}} } c_{\alpha_1, \alpha_2}^{\mu_1, \mu_2} \cdot |\Phi_{\alpha_1, \alpha_2}^{\mu_1, \mu_2}\rangle}_{\rm doubles} + \hdots+
    \underbrace{\sum_{\substack{1\leq \alpha_1 <\hdots <\alpha_{\tilde{N}} \leq N \\ N < \mu_1< \hdots <\mu_{\tilde{N}} \leq N_{\rm tot} }}}_{\tilde{N}-\rm tuples}c^{\mu_1 \hdots \mu_{\tilde{N}}}_{\alpha_1 \hdots \alpha_{\tilde{N}}}|\Phi^{\mu_1 \hdots \mu_{\tilde{N}}}_{\alpha_1 \hdots \alpha_{\tilde{N}}} \rangle.
\end{equation}
Note that in the above expression the coefficients $c_{\alpha_1\hdots\alpha_n}^{\mu_1\hdots\mu_n}$ are given, up to a sign, by the product 
$$
\prod_{1\leq i,j \leq n}\phi_{\alpha_i}(f_j(x)).
$$

\subsection{Concluding remarks}
$\;$

It must be noted that for arbitrary $\tilde{N}$ not all of the coefficients of the above expansion can be independent from each other. This claim is supported by a simple counting argument. There are $\tilde{N}\cdot (\tilde{N} + N)$ of the  $\phi_\alpha^{\rm HF}(f(x))$ coefficients from the bottom hidden sub-matrix. However, in general , the number of free parameters in the above CI expansion is:
$$
\sum_{1\leq i \leq \tilde{N}} \binom{N}{i} \binom{\tilde{N}}{i}, 
$$
which except for $\tilde{N} = 1$ is larger than $\tilde{N}\cdot (\tilde{N} + N)$. While this imposes a constraint on the type of CI wavefunctions that can be obtained with this construction, it must be noted that other successful trial states like unitary coupled clusters also suffer from similar constraints.

While the above construction assumed the staring orbitals to be the occupied and virtual orbitals from the HF solution of the given Hamiltonian, a more general case concerns the choice of arbitrary linear combinations of those orbitals. In  practice we allow such arbitrary linear combinations, and therefore, the trial state we use in this work contains, as a particular case, the CI wavefunction. 

Finally it is important to remark that in this derivation we have set $\phi_\alpha^{\rm HF}(f(x))$ to be constants. However, in the general case they will be bosonic functions of the visible particle positions, which leads to Jastrow-like factors in the CI expansion, thus leading to a more general wave function.

\section{Physically motivated constraint functions}\label{Apndx B: physically motivated}
In this appendix we first show an explicit construction of the Gutzwiller wave function {\it ansatz} from the hidden fermion determinant state. This construction serves as a motivation to showcase the fact that a simple constraint function leads to a correlated wave function. Inspired by this, we present other physically motivated (non-optimizable) constraint functions, and benchmark them in the $4\times 4$ Hubbard model at quarter occupation. We also study the possibility of adding neural-network-based correlation factors to the hidden fermion Slater determinant state.
\subsection{Explicit construction of the Gutzwiller wave function in the Slater determinant hidden fermion state}
In order to gain intuition about the nature of the hidden fermion determinant state, consider as a trivial warmup, the molecular orbital wave function, which has the following Fock-space representation corresponding to $N=2$ fermions in $M = 4$ modes,
\begin{equation}
    \left[c_{1\uparrow}^\dag c_{2\downarrow}^\dag 
    + 
    c_{2\uparrow}^\dag c_{1\downarrow}^\dag 
    +
    e^{-g}(c_{1\uparrow}^\dag c_{1\downarrow}^\dag
    +
    c_{2\uparrow}^\dag c_{2\downarrow}^\dag)\right] |0\rangle \enspace .
\end{equation}
by introducing one hidden fermion ($\tilde{N} = 1$) with two hidden states ($\tilde{M} = 2$) and using the following disjointly supported orbitals,
\begin{align}
    \phi_{\rm v}
    & =
    \begin{bmatrix}
    \phi_1(1{\uparrow}) & 0 \\
    \phi_1(2{\uparrow}) & 0 \\
    0 & \phi_2(1{\downarrow})\\
    0 & \phi_2(2{\downarrow})
    \end{bmatrix}
    \enspace ,
    & 
    \chi_{\rm v}
    =
    \begin{bmatrix}
    0 \\
    0 \\
    0 \\
    0
    \end{bmatrix}
    \enspace , \\
    \phi_{\rm h}
    & =
    \begin{bmatrix}
    0 & 0 \\
    0 & 0
    \end{bmatrix} \enspace ,
    &
    \chi_{\rm h} =
    \begin{bmatrix}
    1 \\
    e^{-g}
    \end{bmatrix}
    \enspace .
\end{align}
In particular, with the following choice of constraint function
\begin{equation}
    f(x) = n_{1\uparrow}n_{1\downarrow}+n_{2\uparrow}n_{2\downarrow} \enspace ,
\end{equation}
we find the probability amplitudes
\begin{align}
\psi(x)
    & =
    \det
    \begin{bmatrix}
    \phi_1(i_1) & \phi_2(i_1) & 0 \\
    \phi_1(i_2) & \phi_2(i_2) & 0 \\
    0 & 0 & \exp\left[-g\sum_{i=1}^2 n_{i\uparrow}n_{i\downarrow}\right]
    \end{bmatrix}
    \\
    & = 
    \exp\left[-g\sum_{i=1}^2 n_{i\uparrow}n_{i\downarrow}\right]
    \det
    \begin{bmatrix}
    \phi_1(i_1) & \phi_2(i_1) \\
    \phi_1(i_2) & \phi_2(i_2)
    \end{bmatrix} \enspace ,
\end{align}
which is of the required form. By similar reasoning, one can construct the Gutzwiller projection of an arbitrary Slater determinant state with $2M$ modes described by orbitals $\phi_{\rm v}$,
\begin{equation}
    \psi(x) = \exp\left[-g\sum_{i=1}^M n_{i\uparrow}n_{i\downarrow}\right] \det \big[ \phi_{\rm v}(x) \big] \enspace ,
\end{equation}
simply by choosing disjointly supported hidden and visible orbitals $\phi_{\rm h} = 0$, $\chi_{\rm v} = 0$, $\chi_{\rm h} = \big[1, e^{-g},\ldots,e^{-Mg}\big]^{\rm T}$ together with the constraint function
\begin{equation}
    f(x)= \sum_{i=1}^M n_{i\uparrow}n_{i\downarrow} \enspace .
\end{equation}

\subsection{Benchmarks with physically motivated constraint functions}

\begin{figure*}
            \includegraphics[width=.99\linewidth]{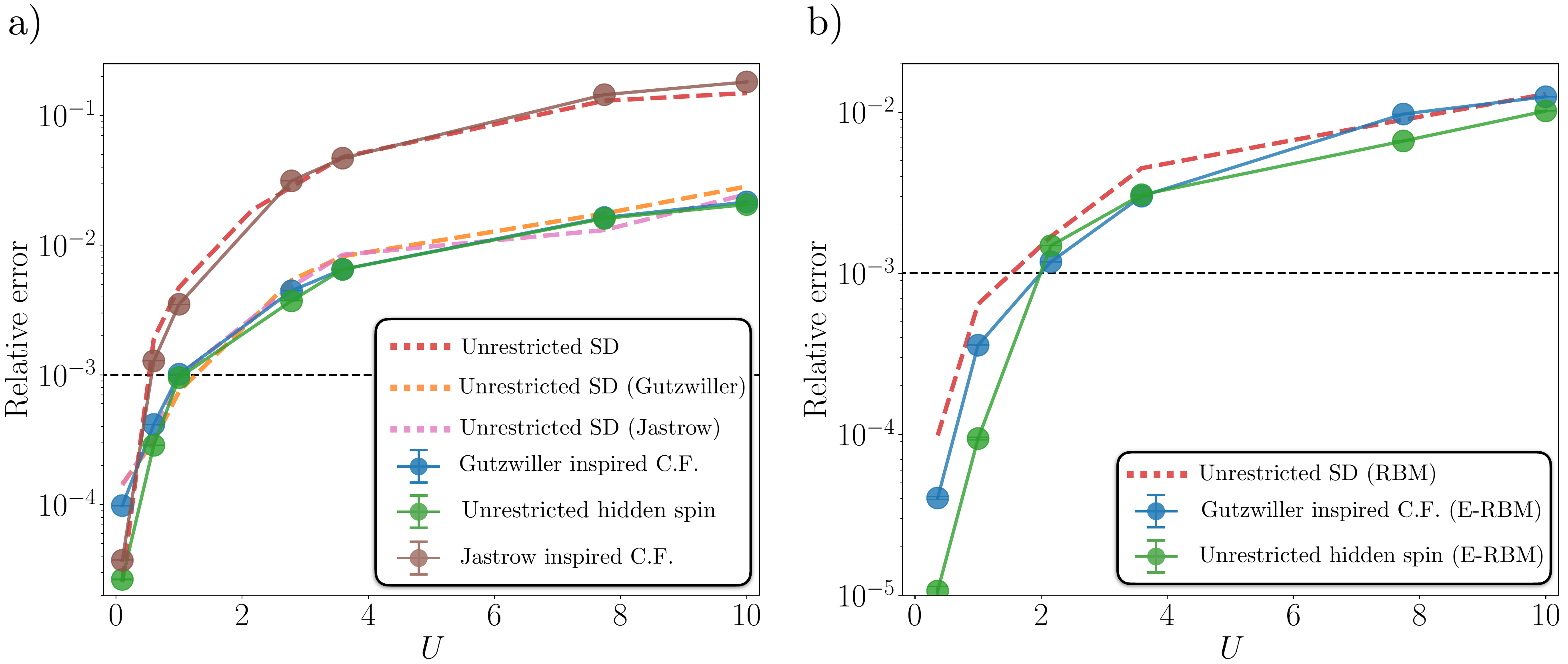}
            \caption{\label{fig_A_01: phys motivated} Benchmarks of physically motivated constraint functions with ED energies in the $4\times 4$ Hubbard model at $n = 1/2$ average physical site occupation. Results from standard wave function {\it ans\"{a}tze}  are shown as dashed lines for comparison purposes. (a) Relative error in the ground-state energy as a function of the coupling constant $U$. The different constraint function {\it ans\"{a}tze}  are a single Slater determinant in the augmented Fock space with no projections. (b) Same as panel (a) including a complex RBM projection factor both in the control unrestricted HF {\it ansatz} and a E-RBM factor in the the hidden fermion {\it ans\"{a}tze}.}
\end{figure*} 
Motivated by the ability to reproduce correlated wave function {\it ans\"{a}tze}  as described above, we propose and test a collection of constraint functions built from the physical intuition about the ground state physics of the Hamiltonian in Eq.~(1) in the main text.

    \subsubsection{Gutzwiller-inspired constraint function} Take $\tilde{N} = 1$ and $\tilde{M} = N/2$, with the constraint function $f(x)= \sum_{i=1}^M n_{i\uparrow}n_{i\downarrow}$. The position of the hidden fermion in the hidden modes is determined by the number of doubly occupied sites in the $x$ configuration. The wave function {\it ansatz} is the Slater determinant in the enlarged Hilbert space of Eq.~(5) in the main text. The variational parameters are the coefficients (orbitals) of the non-orthogonal change of single-particle basis in Eq.~(6) in the main text. For a particular choice of orbitals this {\it ansatz} can reproduce the Gutzwiller wave function as described above.
    
    \subsubsection{Unrestricted hidden spin {\it ansatz}} Consider a pair of hidden fermionic modes $\hat{d}^\dag_{i\beta} $ with $ \beta = 0, 1$, associated to each site of the visible Hubbard lattice $i$, and a single hidden fermion populating each pair of hidden modes. Thus $\tilde{N} = M/2$ and $\tilde{M} = M$. The position of the hidden fermion depends on the visible site occupancy via the local constraint function $f_i(x) = n_{i\uparrow}n_{i\downarrow} = \beta$. The wave function {\it ansatz} is the Slater determinant in the enlarged Hilbert space of Eq.~(5) in the main text. The variational parameters are the coefficients (orbitals) of the non-orthogonal change of single-particle basis in Eq.~(6) in the main text. This {\it ansatz} is closely related to the hidden spin formalism~\cite{Medici2005slavespins}. The difference is to consider the hidden fermions to be indistinguishable particles amongst each other and the visible ones, as opposed to the hidden spin formalism where the added spins are distinguishable particles. This unrestricted hidden spin {\it ansatz} provides more flexibility, the same way the unrestricted Hartree-Fock does compared to factorised HF. It is easy to see that the unrestricted hidden spin {\it ansatz} can reproduce the Gutzwiller {\it ansatz} for a particular choice of orbitals. Furthermore, it was proven that the hidden spin formalism can capture the Mott transition at the mean field level. Therefore, this {\it ansatz} can capture correlations beyond the Gutzwiller single-site correlations.

    \subsubsection{Jastrow-inspired constraint function} Consider a disjointed collection of hidden modes $\hat{d}^\dag_{l, \beta}$ with $1\leq l \leq D_{\rm max}$ and $0 \leq \beta \leq N$, where $D_{\rm max}$ is the maximum graph distance (Manhattan distance) in the physical lattice. There is one hidden fermion for each of the $l$ groups of hidden modes, $\tilde{N} = D_{\rm max}$. The position of the hidden particle at the $l^{\rm th}$ group of hidden modes depends on the number of visible fermions at distance $l$ in the $x$ configuration: 
    \begin{equation}
     f_l(x)  = \left\lfloor \frac{1}{\tilde{N}} \sum_{i \in \mathcal{V}} \sum_{j \in (i+l)} n_in_j. \right\rfloor
    \end{equation}
    
    \subsubsection{Benchmarks}
    Panel (a) on Fig.~\ref{fig_A_01: phys motivated} shows the relative error in the ground state energy as a function of $U$ for the  Gutzwiller-inspired constraint function, unrestricted hidden spin and Jastrow-inspired constraint function {\it ans\"{a}tze}. It also shows the relative error for standard wave function {\it ans\"{a}tze}  like the unrestricted SD and the unrestricted SD with Gutzwiller and Jastrow factors for comparison. The Gutzwiller-inspired constraint function, and unrestricted hidden spin {\it ans\"{a}tze}  have a relative error comparable to it of the standard unrestricted SD with Gutzwiller and Jastrow factors, providing for some values of $U$ a marginal improvement in the ground-state energy. On the other hand, the Jastrow-inspired constraint function {\it ansatz} achieves relative error values comparable to a single unrestricted SD. These results show that the choice of constraint function is of great relevance to find and {\it ansatz} that can capture correlated phenomena. 

    In addition to considering mean-field states in the enlarged Hilbert space, correlated states can also be constructed by multiplying the Slater determinant by a parametrized factor that depends on both the occupation of the visible and hidden degrees of freedom. In particular we take that factor to be a complex Restricted Boltzmann Machine in the enlarged space of occupations $n \oplus \tilde{n}$ (E-RBM). Panel (b) of Fig.~\ref{fig_A_01: phys motivated} shows the relative error in the ground-state energy as a function of the coupling constant for the  The Gutzwiller-inspired constraint function, and unrestricted hidden spin {\it ans\"{a}tze}  with an additional E-RBM factor where the hidden layer has twice as many units as input features ($\alpha = 2$). For comparison purposes, we also show the relative error of an unrestricted SD {\it ansatz} (visible space) with an additional complex RBM factor with twice as many units as input features ($\alpha = 2$), proposed proposed for the first time in Ref.~\cite{Nomura2017slaterRBM}. At smaller values of $U$ the correlated {\it ans\"{a}tze}  in the enlarged space have a lower relative error in the energy than the equivalent correlated {\it ansatz} in the original Hilbert space. At larger values of $U$, the correlated unrestricted hidden spin {\it ansatz} notably outperforms the correlated {\it ansatz} in the original Hilbert space.
    
    While the results presented above show that the hidden fermion formalism outperforms well established wave functions {\it ans\"{a}tze}  for carefully chosen constraint functions, it is still unclear how to choose the optimal constraint function. In fact, there exists a doubly exponential number of constraint functions, making the brute-force search of constraint function an intractable problem.

\section{Effect of the number of hidden fermions and neural network architecture in the expressive power of the hidden fermion determinant state} \label{Apndx D: depthd and N_hidden}
  \begin{figure*}
            \includegraphics[width=.99\linewidth]{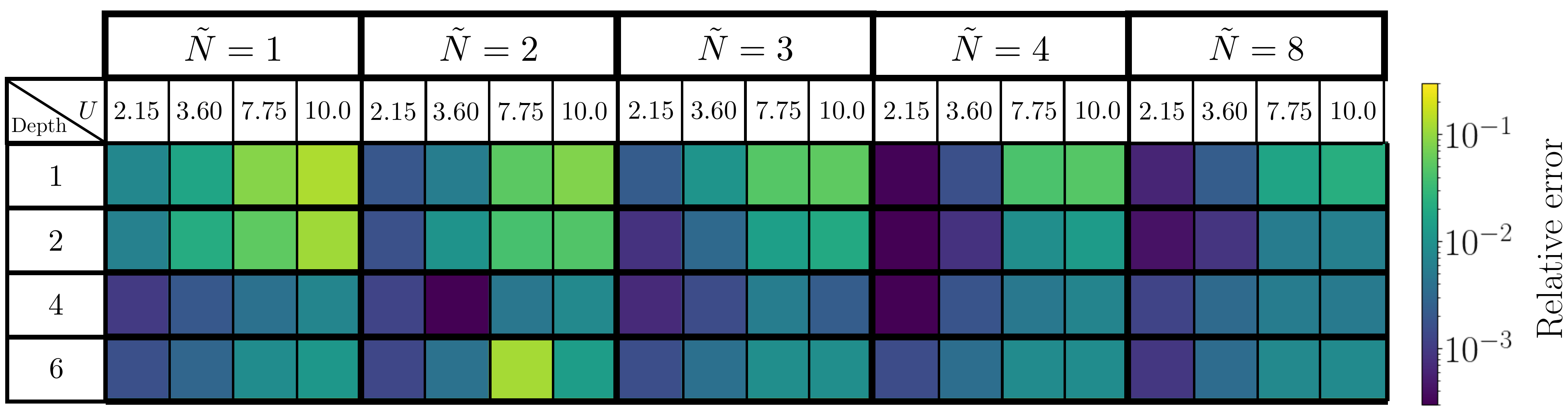}
            \caption{\label{fig_A_02: Nh and depth effect} Effect of the number of hidden fermions and depth of the fully connected neural network that parametrizes the hidden sub-matrix in the expressive power of the hidden fermion determinant {\it ansatz}. The color scale shows the relative error in the ground-state energy for different values of $U$ given $N_h$ and the neural network depth. The results correspond to the  $4\times 4$ Hubbard model at $n = 1/2$ filling.}
\end{figure*} 
We study the effect of the depth of the neural networks, leaving the width of the hidden layers fixed. The width of the first half of hidden layers matches the number of input features and the second half matches the number of output features. Fig.~\ref{fig_A_02: Nh and depth effect} shows the relative error in the ground-state energy for different values of $U$ and $\tilde{N}$ and the neural network depth. For every number of hidden fermions, the accuracy improves as the depth of the neural network increases until a critical value is reached, in this case six hidden layers. Beyond six hidden layers the optimization becomes challenging . Even though deeper architectures are known to be more expressive\cite{montufar2014deepernets}, the optimization of the energy becomes increasingly challenging, making the {\it ansatz} to get stuck in a local minimum that can be identified with the mean-field solution. It is also noteworthy that for a fixed depth, the error decreases as the number of hidden fermions is increased, to the point that shallow architectures lead to low errors. 

In particular, note that a neural network with a single hidden layer (depth 2) and with eight hidden fermions provides the lowest error for the whole range of coupling strengths. We choose this configuration of eight hidden fermions and a single hidden layer architecture as a starting point to study the effect of the width of the network in the {\it ansatz} expressive power in the main text. The key advantage is that in this network architecture, the number of hidden units is the only control parameter to increase its expressivity. Furthermore, this single hidden layer architecture is the minimal architecture that satisfies the universal approximation theorem\cite{Cybenko1989universal}. 

\section{Energy-variance extrapolation in Fig. 3 (a)} \label{Apndx E: variance extrapolation}
The convergence of the resulting ground-state energy with the complexity of the neural-networks in the variational states is not a well understood matter, thus, the direct extrapolation of the energy from the parameter controlling the NN complexity is not a well motivated technique. However, it is clear that in general more expressive architectures yield lower energies and variances. This collection of converged energies and variances can be used to extrapolate a better estimate to the ground-state energy. 

The energy-variance extrapolation is motivated by the fact that the difference between the true ground-state energy $\langle \hat{H}\rangle_{\rm gs}$ and the variational expectation value $\langle \hat{H} \rangle$:
\begin{equation}
    \delta E = \langle \hat{H} \rangle-\langle \hat{H}\rangle_{\rm gs}
\end{equation}
vanishes linearly with the energy variance
\begin{equation}
    \Delta{E} = \frac{\langle \hat{H}^2 \rangle-\langle \hat{H} \rangle^2}{\langle \hat{H} \rangle^2}
\end{equation}
for sufficiently small $\delta E$ (see Eqs. 2.17-2.23. in Ref~\cite{Kashima2001varianceExtrapolation}). Therefore, a better estimate to $\langle \hat{H}\rangle_{\rm gs}$ can be extracted by performing linear regression on the variational energies and variances coming from different trial wave function with different NN complexities. In our case for different values of the ratio between hidden and input units $\alpha$.

\begin{figure*}
            \includegraphics[width=.99\linewidth]{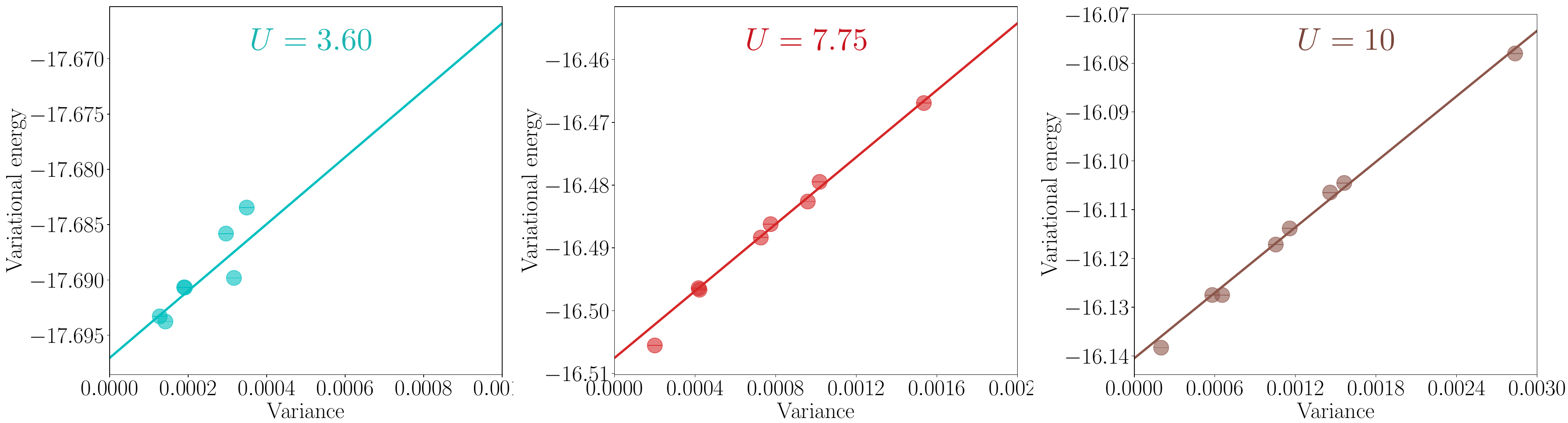}
            \caption{\label{fig_A_03: variance extrapolation} Energy-variance extrapolation of the ground-state energy for the $4 \times 4$ square Hubbard lattice with periodic boundary conditions and average site occupation $n = 1/2$. The variational energy for different values of the ratio between hidden and input units $\alpha$ is shown as a function of its variance. The data is the same as in Fig.~3 (a) in the main text. Different panels show the the extrapolation for different values of the coupling constant $U$. The solid lines are the lines of best fit to the variational data. }
\end{figure*}

The variational energies from Fig.~3 (a) in the main text for different values of $\alpha$ are shown as a function of their variance in Fig.~\ref{fig_A_03: variance extrapolation}, at different values of the coupling constant $U$. The relation between the variational energy and the variance is in good agreement with a linear trend. From the intercept of the linear extrapolation with the zero variance axis we extract estimates of $\langle \hat{H}\rangle_{\rm gs}$. The relative errors of the extrapolated energies are shown in Fig.~3 (a) in the main text.

\section{Energy benchmarks in the $L\times L$ Hubbard model at half filling.}
We study the accuracy, of the hidden fermion determinant state, relative to Auxiliary Field Quantum Monte Carlo (AFQMC), on increasingly larger systems. At half filling AFQMC does not have a sign problem. Therefore it provides very accurate estimates of the energy values.

\begin{figure}
    \centering
    \includegraphics[width=.45\linewidth]{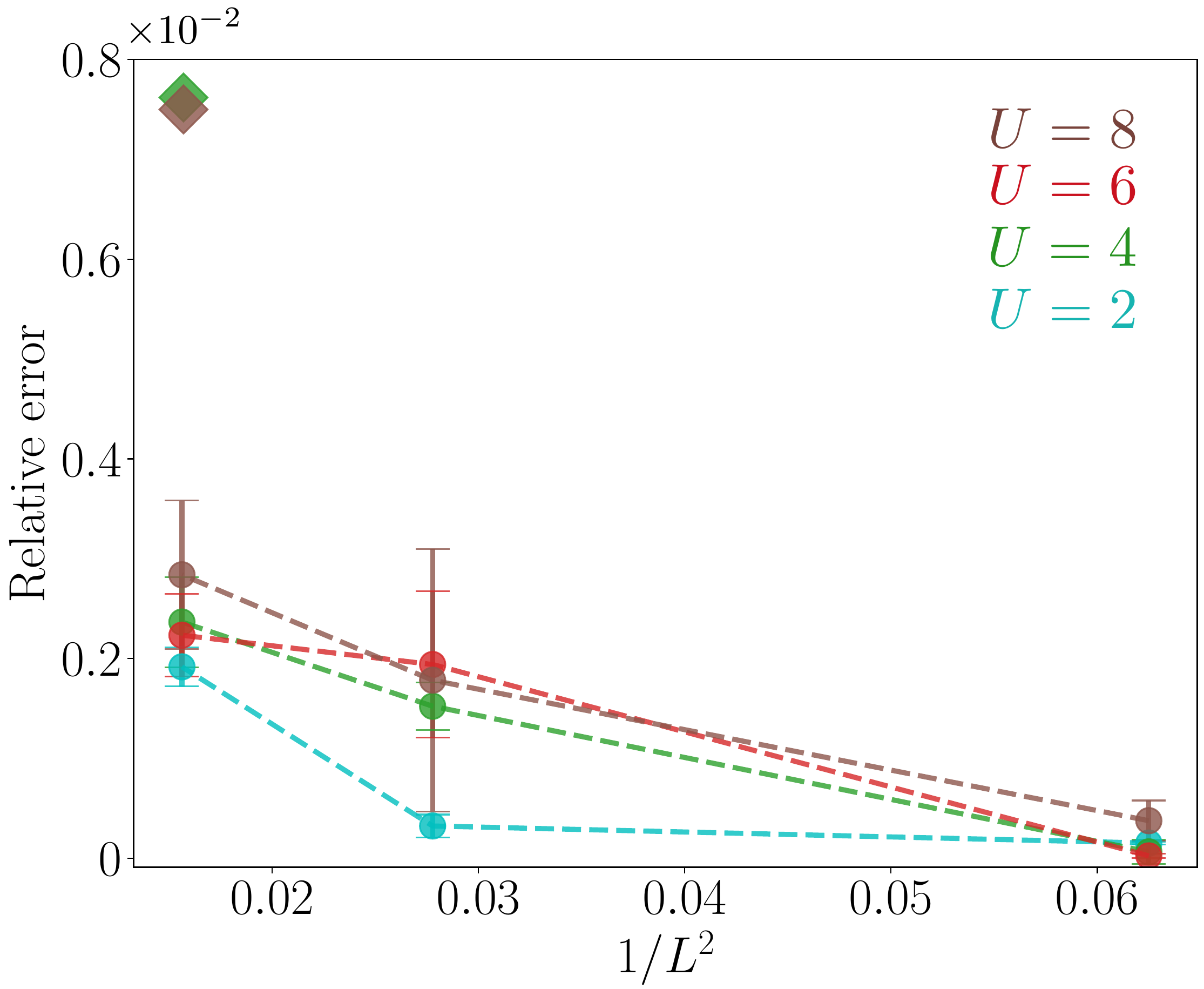}
    \caption{\label{fig_A_04: benchmark at half filling} Relative (to AFQMC energies from Ref.~\cite{Qin2026AFQMCBenchmark}) error in the ground-state energy in the square lattice Hubbard model at half filling for increasing side length $L$. Periodic boundary conditions are imposed along one of the sides while anti-periodic boundary conditions are imposed along the other side. Different values of $U$ are shown in different colors as indicated. The green and brown diamonds correspond to the accuracy reported in Ref.~\cite{Nomura2017slaterRBM} using a neural-network based Jastrow wave function ansatz for $U = 4$ and $U = 8$ respectively. }
\end{figure}

We focus on the $4\times 4$, $6\times 6$ and $8\times 8$ system sizes, imposing periodic boundary conditions along one of the sides of the square and anti-periodic along the other side.  In this case the energy per site increases monotonically with the side length~\cite{Qin2026AFQMCBenchmark}. We use the hidden fermion determinant state with a fully parametrized hidden sub-matrix. The number of hidden fermions and hidden unit densities are $\tilde{N} = \{8, 8, 16 \}$ and $\alpha = \{ 96, 78, 1\}$ for the $4\times 4$, $6\times 6$ and $8\times 8$ lattices respectively. Figure~\ref{fig_A_04: benchmark at half filling} shows the relative error in the ground-state energy as a function of the inverse of the number of sites in the lattice for different values of $U$. While the error increases with the system size, it always stays in the order $\mathcal{O}\left(10^{-3}\right)$ or lower. Remarkably, the achieved accuracy is better, by at least a factor of two, than the reported accuracy for the neural-network based wave function ansatz in Ref.~\cite{Nomura2017slaterRBM}.

\section{First quantized fermions on the open-sourced library NetKet: a VMC library for distinguishable particles}\label{Apndx F: Netket}
Recall that in the canonical quantization of $N$ spinless fermions hopping on an undirected graph $G=(\mathcal{V},\mathcal{E})$, the configuration space is described by arrays of the form $x = (x_1,\ldots,x_N) \in \mathcal{V}^N$, where we assume some ordering on the vertices, which identifies them with the integers from $1$ to $|\mathcal{V}|$. In this section we describe how to reformulate the problem in terms of a local spin model necessary for implementation in the NetKet framekwork. The Hilbert space of the spin model consists of a tensor product of local Hilbert spaces $\mathbb{C}^{N+1}$
\begin{equation}
    \mathcal{H}_{\rm spin} = \bigotimes_{i \in \mathcal{V}} \mathbb{C}^{N+1}
\end{equation}
which is spanned by orthonormal basis vectors of the form $|\kappa\rangle$ where $\kappa : \mathcal{V} \to \{0,\ldots,N\}$. Recall that each function $\kappa$ provides lookup table describing the location of a vertex within the configuration $x = (x_1,\ldots,x_N)$ (or zero if the vertex is absent). It is clear that $\mathcal{H}_{\rm spin}$ contains many unphysical states. These unphysical states are avoided in the Markov chain by ensuring a valid initialization and using a transition rule that exchanges the state of two vertices. Now we discuss the implementation of the following operators as local operators on $\mathcal{H}_{\rm spin}$,
\begin{equation}
    n_i \enspace , \quad \quad n_in_j \enspace, \quad \quad c_i^\dag c_j
\end{equation}
The first is represented by a single-site operator $ n:\mathbb{C}^{N+1} \to \mathbb{C}^{N+1}$ given by $n := \mathbbm{1} - |0 \rangle \langle 0 |$
where $\mathbbm{1}$ denotes the identity operator on $\mathbb{C}^{N+1}$. The second is given by the tensor product operator $n \otimes n$, where the left and right tensor factors represent sites $i$ and $j$, respectively. Finally, $c_i^\dag c_j$ is represented by $\big(|0 \rangle \langle 0| \otimes n\big) \mathrm{SWAP}$.

\section{Converged values of variational energies for different trial wave functions}\label{Apndx G: Energies}
This section contains the numerical values of the converged ground state energies obtained in this work, with special emphasis on the hidden fermion determinant state (HFDS). We also include energies from other trial states for reference. Exact diagonalization energies used in the benchmarks are also provided for the smaller system sizes. 

\subsection{$4\times 4$ Hubbard model with $n = 1/2$}
$\;$\newline

Here we provide the variational and exact diagonalization (ED) energies for different trial states in the square lattice Hubbard model of size $4\times 4$ and average site occupation $n = 1/2$ and periodic boundary conditions, for various values of $U$.

Table~\ref{tab01} contains the ground-state variational energies for different values of $U$ obtained with the HFDS wave function with a parametrized constraint function. The hidden sub-matrix is parametrized by single-hidden-layer neural networks of width density $\alpha$. It also contains the ED energies for reference. The data corresponds to the relative errors shown in Fig.~3 (a) in the main text.

\begin{table}[h!]
\centering
\begin{tabular}{||c || c | c | c | c | c | c | c | c ||} 
 \hline
 U & ED energy & HFDS ($\alpha = 0.5$) & HFDS ($\alpha = 1$) & HFDS ($\alpha = 2$) & HFDS ($\alpha = 4$) & HFDS ($\alpha = 8$) & HFDS ($\alpha = 16$) & HFDS ($\alpha = 32$)\\ [0.5ex] 
 \hline\hline
 3.6 & -17.6980300(1) & -17.65651(7) & -17.68383(9) & -17.6893(1) & -17.6897(4) & -17.6925(6) & -17.6929(8) & -17.69357(4)\\ 
 \hline
 7.75 &-16.5091548(2) &  -16.4196(4) & -16.4823(6) & -16.4893(3) & -16.4959(4) & -16.4960(3) & -16.502(3) & -16.505(1)\\
 \hline
 10 &  -16.1432081(7) &  -16.0457(1) & -16.0995(5) & -16.1103(8) & -16.1204(5) & -16.1329(4) & -16.1348(5) & -16.1383(9)\\
 \hline
\end{tabular}
\caption{Ground-state energy for different values of $U$ in the $4\times 4$ Hubbard model with $n = 1/2$ and periodic boundary conditions. Table contains ED energies for reference. Table also contains variational energies from the hidden fermion determinant state with a parametrized constraint function. The data corresponds to the relative errors shown in Fig.~3 (a) in the main text. The hidden sub-matrix is parametrized by single-hidden-layer neural networks of width density $\alpha$. See the main text for more details.}\label{tab01}
\end{table}

Table~\ref{tab02} contains ground-state energies for different variational states, with an emphasis on variational energies from hidden fermion determinant states and hidden fermion determinant states with an added RBM factor in the augmented space with physically motivated constraint functions, as described in Section~3 in the Supplementary information. Other trial states are included for reference. It also contains the exact diagonalization (ED) energies used for benchmarking. The data corresponds to the relative errors shown in Fig.~1 in the Supplementary Information.

\begin{table}[h!]
\centering
\begin{tabular}{||c || c | c | c | c | c | c| c ||} 
 \hline
 \textit{Ansatz} & U = 0.1 & U = 0.6& U = 1& U =2.15 & U = 3.6& U = 7.75& U =10\\ [0.5ex] 
 \hline\hline
 ED & -19.9073095(9) & -19.4747633(2) & -19.161986(3) & -18.4053208(8) & -17.698030(1) & -16.509154(8) & -16.509154(8)\\ 
 \hline
 USD & -19.906(7) & -19.43(3) & -19.07(4) & -18.06(6) & -16.85(8) & -14.36(8) & -13.74(7)\\
 \hline
 USD (Gutzwiller) & -19.9044(7) & -19.468(1) & -19.147(2) & -18.306(3) & -17.552(1) & -16.219(1) & -15.685(2)\\
 \hline
 USD (Jastrow) & -19.90449(1) & -19.46832(6) & -19.1432(1) & -18.3173(2) & -17.5487(3) & -16.2932(5) & -15.7443(6) \\
 \hline
 USD (RBM) & -19.904895(4) &  -19.47284(1) &  -19.14965(3) &  -18.37420(7) &  -17.6184(1) & -16.3607(2) &  -15.9339(2) \\
 \hline
 Gutzwiller inspired CF & -19.90535(1) & -19.46667(2) & -19.14257(3) & -18.32346(8) & -17.58269(9) & -16.2395(1) & -15.7953(2) \\
 \hline
 Unrestricted hidden spin & -19.906781(3) &  -19.46918(1) &  -19.14371(2) &  -18.33683(7) &  -17.58288(8) &  -15.8113(1) &  -16.1697(2) \\
 \hline
 Jastrow inspired CF & -19.906567(7) &  -19.44978(4) &  -19.09486(7) &  -17.8293(2) &  -16.8694(2) &  -14.1253(4) &  -13.2197(5)  \\
 \hline
 Gutzwiller inspired CF(RBM) & -19.904895(4) & -19.47284(1) & -19.14965(3) & -18.37420(6) & -17.6184(2) & -16.3607(2) & -15.9416(2) \\
 \hline
 Unrestricted hid. spin(RBM) & -19.904427(5) &  -19.47455(1) & -19.16018(3) &  -18.37803(5) & -17.64362(9) & -16.3997(2) & -15.9788(2) \\
 \hline
\end{tabular}
\caption{Ground-state energy for different values of $U$ in the $4\times 4$ Hubbard model with $n = 1/2$ and periodic boundary conditions. Table contains ED energies for reference. The table contains variational energies for various trial states, including unrestricted Slater determinants (USD), unrestricted Slater determinants with applied Gutzwiller factor (USD Gutzwiller), unrestricted Slater determinants with applied Jastrow factor (USD Jastrow), unrestricted Slater determinants with applied RBM factor (USD RBM), as well as HFSD with physically motivated constraint functions: Gutzwiller inspired CF, unrestricted hidden spin and Jastrow inspired CF (see Section 3.~B in the Supplementary Information for details). The table also contains results for the HFSD state with an added RBM factor in the augmented space, using the Gutzwiller inspired and unrestricted hidden spin constraint functions (see Section 3.~B in the Supplementary Information for details). The data corresponds to the relative errors shown in Fig.~1 in the Supplementary Information. 
}\label{tab02}
\end{table}

\subsection{$4\times 4$ Hubbard model with $n = 5/8$}
$\;$\newline

Table~\ref{tab03} provides the variational (for the hidden fermion determinant state with a fully parametrized hidden sub-matrix) and exact diagonalization (ED) energies in the square lattice Hubbard model of size $4\times 4$ and average site occupation $n = 5/8$ (first closed shell of the model) and periodic boundary conditions, for various values of $U$. The hidden sub-matrix is parametrized by single-hidden-layer neural networks of width density $\alpha$. Projections to different symmetry subspaces are applied to the converged trial state: rotation of $\pi/2$ ($C_4$), translations of momentum $K = 0$ and the intersection of both subspaces. See the main text for more details. The data corresponds to the relative errors displayed in Fig.~3 (b) in the main text.

\begin{table}[h!]
\centering
\begin{tabular}{||c ||  c | c | c| c ||} 
 \hline
 \textit{Ansatz} & U =2.15 & U = 3.6& U = 7.75& U =10\\ [0.5ex] 
 \hline\hline
 ED & -21.2122357(7) & -19.8916371(9) & -17.6037315(5) & -16.9035599(8)\\ 
 \hline
 HFSD ($\alpha = 64$) & -21.211411(4) & -19.888814(8) & -17.60017(2) & -16.90020(3)\\ 
 \hline
 HFSD ($\alpha = 64$)+$C_4$ & -21.211811(4) & -19.889250(7) & -17.60100(2) & -16.90119(2)\\ 
 \hline
 HFSD ($\alpha = 64$) + $K = 0$ & -21.211853(3) & -19.890244(6) & -17.60140(2) & -16.90147(2)\\ 
 \hline
 HFSD ($\alpha = 64$) + ($C_4 \cap K = 0$) & -21.211896(2) & -19.890443(6) & -17.60200(1) & -16.90183(2)\\ 
 \hline
\end{tabular}
\caption{Ground state energy for different values of $U$ in the $4\times 4$ Hubbard model with $n = 5/8$ and periodic boundary conditions. Table contains ED energies for reference. The table shows the converged energies for the hidden fermion determinant state with a fully parametrized hidden sub-matrix. Energies of the projection of the converged trial state to different symmetry subspaces are also shown. The data corresponds to the relative errors displayed in Fig.~3 (b) in the main text.}\label{tab03}
\end{table}

\subsection{$4\times L$ Hubbard model with $n = 7/8$ and U = 8}
$\;$\newline

Here we present the converged ground-state energies of the hidden fermion determinant state with a fully parametrized hidden sub-matrix in the Hubbard model in rectangular geometries of dimension $4\times L$ at $1/8$ hole doping ($n = 7/8$) and $U = 8$. Periodic boundary conditions are considered along the short side of the rectangle, while both open and periodic boundary conditions are taken along the long side of the rectangle. In Table~\ref{tab04} both configurations are labelled as PBC-PBC and PBC-OBC respectively. Table~\ref{tab04} shows the variational energy per site obtained with the hidden fermion determinant state with a fully parametrized hidden sub-matrix. The value of $\alpha$ depends on the system size (see Section 3.~B in the main text for details). The energies are those displayed in Fig.~4 in the main text.

\begin{table}[h!]
\centering
\begin{tabular}{||c || c |c ||} 
 \hline
 L & Energy per site HFDS (PBC-PBC)& Energy per site HFDS (PBC-OBC)\\ [0.5ex] 
 \hline\hline
 4 & -0.7409(1) & NA\\ 
 \hline
 8 & -0.7633(7) & -0.7309(6) \\ 
 \hline
 16 & -0.753(2) & -0.7466(8)\\ 
 \hline
\end{tabular}
\caption{ Variational energy per site in the Hubbard model in rectangular geometries of dimensions $4\times L$ at $1/8$ hole doping ($n = 7/8$) and $U = 8$, obtained with the hidden fermion determinant state with a fully parametrized hidden sub-matrix. The value of $\alpha$ depends on the system size (see Section 3.~B in the main text for details). The energies are those displayed in Fig.~4 in the main text.}\label{tab04}
\end{table}

\subsection{$L\times L$ Hubbard model with $n = 1$}
$\;$\newline

Table~\ref{tab05} shows the energy per site on the square lattice Hubbard model of size $L\times L$ with periodic boundary conditions along one of the sides of the square and anti-periodic boundary conditions along the other side. The trial state is the hidden fermion determinant state with a fully parametrized hidden sub-matrix. Details on the number of hidden fermions and hidden unit densities can be found on Section~6 of the Supplementary Information.

\begin{table}[h!]
\centering
\begin{tabular}{||c || c |c |c |c ||} 
 \hline
 L & U = 2 & U = 4 & U = 6 & U = 8\\ [0.5ex] 
 \hline\hline
 4 & -1.25693(1) & -0.9120(1) & -0.68135(1) & -0.5401(1)\\ 
 \hline
 6 & -1.2079(1) & -0.8717(2) &  -0.6609(4) & -0.5289(3)\\ 
 \hline
 8 & -1.1900(2) & -0.8621(4) &  -0.6574(2) & -0.5244(6)\\ 
 \hline
\end{tabular}
\caption{ Variational energy per site in the $L\times L$ Hubbard model at half filling with periodic boundary conditions  along one of the sides of the square and anti-periodic boundary conditions along the other side. The trial state is the hidden fermion determinant state with a fully parametrized hidden sub-matrix (see Section~6 in the Supplementary Information for more details). The energies correspond to the relative errors shown in Fig~4 in the Supplementary Information. }\label{tab05}
\end{table}


\end{document}